\documentclass[12pt]{article}
\usepackage{amsmath}
\usepackage{graphicx,psfrag,epsf}
\usepackage{enumerate}
\usepackage{natbib}
\usepackage{url} % not crucial - just used below for the URL 

\usepackage{dsfont,blkarray,bm,amssymb,amsfonts,mathrsfs,amsthm}
\usepackage{breqn}
\usepackage{pifont}
\usepackage{bbm}
\usepackage[makeroom]{cancel}
\usepackage[table]{xcolor}
\usepackage[hidelinks]{hyperref}
\usepackage{multirow}
\hypersetup{
linkcolor=blue,
citecolor=blue,
urlcolor=blue,
colorlinks=true
}
\usepackage{tabularx,booktabs}
\usepackage{scalerel}
\usepackage{algorithm}
\usepackage{algcompatible}
\usepackage{subfigure}
\usepackage{float}
\usepackage{diagbox}
\usepackage{rotating}
\usepackage{changepage}
\usepackage{graphics}
\usepackage{enumitem}
\usepackage{tablefootnote}
\usepackage{ragged2e}
\usepackage{chngcntr}
\usepackage{setspace}
\usepackage{cases}
\usepackage{arydshln}
\usepackage{pdflscape}
\usepackage{rotating}
\usepackage{changepage}
\usepackage{caption}  % For customizing captions

\newcommand{\blind}{0}

\begin{document}

\def\spacingset#1{\renewcommand{\baselinestretch}%
{#1}\small\normalsize} \spacingset{1}

%%%%%%%%%%%%%%%%%%%%%%%%%%%%%%%%%%%%%%%%%%%%%%%%%%%%%%%%%%%%%%%%%%%%%%%%%%%%%%

\if0\blind
{
  \title{\bf Using Forests in Multivariate Regression Discontinuity Designs}
  \author{Yiqi Liu\thanks{ Corresponding author; 404 Uris Hall, Ithaca NY 14850. We thank Francesca Molinari for her guidance and support throughout this project. We also thank Levon Barseghyan, Yanqin Fan, Douglas Miller, José Luis Montiel Olea, Chen Qiu, Hongyuan Xia, Sen Zeng, and seminar participants at Cornell University and the University of Washington for valuable discussion and helpful comments. All data and replication files can be accessed at \href{https://github.com/yqi3/GRF-RD}{\texttt{github.com/yqi3/RDForest}}.}\hspace{.2cm}\\
  \href{mailto:yl3467@cornell.edu}{\texttt{yl3467@cornell.edu}}\\
    Department of Economics, Cornell University\\
    and \\
    Yuan Qi\\
    \href{mailto:ayqi@uw.edu}{\texttt{ayqi@uw.edu}}\\
    Department of Economics, University of Washington}
  \maketitle
} \fi

\if1\blind
{
  \begin{center}
    {\LARGE\bf Title}
\end{center}
} \fi
\vspace{-0.5cm}
\begin{abstract}
We discuss estimation and inference of conditional treatment effects in regression discontinuity (RD) designs with multiple scores. In addition to local linear regressions and the minimax-optimal estimator more recently proposed by \citet{IW19}, we argue that two variants of random forests, honest regression forests and local linear forests, should be added to the toolkit of applied researchers working with multivariate RD designs; their validity follows from results in \citet{WA18} and \citet{FTAW20}. We design a systematic Monte Carlo study with data generating processes built both from functional forms that we specify and from Wasserstein Generative Adversarial Networks that closely mimic the observed data. We find no single estimator dominates across all specifications: (i) local linear regressions perform well in univariate settings, but the common practice of reducing multivariate scores to a univariate one can incur undercoverage, possibly due to vanishing density at the transformed cutoff; (ii) good performance of the minimax-optimal estimator depends on accurate estimation of a nuisance parameter and its current implementation only accepts up to two scores; (iii) forest-based estimators are not designed for estimation at boundary points and are susceptible to finite-sample bias, but their flexibility in modeling multivariate scores opens the door to a wide range of empirical applications, as illustrated by an empirical study of COVID-19 hospital funding with three eligibility criteria.

\vspace{.5cm}
\vspace{.5cm}
\end{abstract}
\vfill

\newpage
\spacingset{1.45}
\section{Introduction}
\label{sec:intro}
Regression discontinuity (RD) designs are widely used to estimate causal effects in observational settings, where individuals receive a binary treatment if an observed running variable is above some known threshold. Under the potential outcomes framework, \citet{HTV01} show that the treatment is as good as random in the subpopulation arbitrarily close to the threshold, if the conditional expectations of the two potential outcomes are continuous at the cutoff.\footnote{See, e.g., \cite{IL08} and \cite{CT22}, for reviews on RD designs.} Estimation and inference of treatment effects in this context are, therefore, a local problem in the neighborhood of the threshold. Following \citet{HTV01} and \citet{P03}, local polynomial regressions remain the most popular estimation strategy in the empirical RD literature. At the heart of these local methods is bandwidth selection for the weighting kernel. In designs with one running variable, \citet{IK12} propose an optimal bandwidth from minimizing the asymptotic mean squared error (MSE) at the cutoff. This procedure, however, leads to an asymptotically non-vanishing bias. \citet{CCT14} put forth a bias-corrected estimator that first subtracts an estimated bias term and then adjusts the standard error accordingly.

In many situations, the treatment status is determined by more than one criterion. Yet, univariate local linear regressions do not readily generalize to the multivariate case: kernel-based methods are less useful when the dimension far exceeds two \citep[see, e.g., Chapter 6 of][]{HTF16}, and standard bias correction quickly becomes algebraically involved and computationally difficult. Existing methods for multiple scores include multiple bandwidths selected from cross-validation \citep{PWM11, CL18}, and  \citet{Z12} extends the MSE-optimal bandwidth of \citet{IK12} to the bivariate case. But, as pointed out by \citet{IW19}, these methods yield oversmoothing bandwidths and asymptotically biased estimates;\footnote{\citet{CL18} also consider rule-of-thumb bandwidths that scale with $n^{-1/6}$, which are again oversmoothing in the sense that $nh^5\not\to0$.} to avoid selecting a multivariate bandwidth, following the literature on minimax linear estimation \citep[e.g.,][]{armstrong2018optimal, kolesar2018inference}, they propose an estimator that is minimax-optimal over a class of conditional expectations with second derivatives bounded by a given constant $\mathcal{B}$. This method, however, requires estimating $\mathcal{B}$ and, as we show in Section \ref{sec:simulation}, its performance depends crucially on accurate estimation of this nuisance parameter.

In this paper, we conduct a systematic Monte Carlo simulation to evaluate existing approaches to multivariate RD designs, including the minimax-optimal estimator of \citet{IW19} and the commonly used local linear regression approach with a transformed univariate score: to sidestep the complications of running local linear regressions in higher dimensions, empirical researchers often reduce the problem to a univariate one. Our contribution is twofold. First, we point out a potential issue of the common practice of implementing univariate local linear regressions when the running variable is in fact multivariate, give a formal characterization of when such problem arises, and show in simulations that it can produce biased estimates even for very large sample sizes. Second, we demonstrate two additional valid, easy-to-implement estimators that can be leveraged in multivariate RD designs, and together with a systematic Monte Carlo evaluation of existing estimators and an empirical application, provide researchers with a menu of nonparametric methods that they can tailor to different empirical settings.

\subsection{Related Literature}
Despite the prevalence of multi-score RD designs, empirical researchers often adopt a univariate approach. For example, unemployment insurance programs usually set multiple eligibility criteria, such as age \textit{and} work history in determining the maximum duration of benefits, but \citet{SVB12} subset the data to consider only individuals with eligible employment histories within their respective age groups, and use age as the effective running variable \citep[see also][for another example]{NW17}. A similar subsetting approach is taken by \citet{JL04} and \citet{M08} to evaluate education policies where treatment eligibility is determined by \textit{multiple} test scores. Another common approach is to transform the multivariate score into a univariate one, especially when the treatment boundary takes arbitrary shapes and finding a suitable subset where a univariate score applies is infeasible. For example, to exploit geographic discontinuities, \citet{B99} considers the shortest distance to the boundary of a geographic region; \citet{KT15} transform bivariate coordinates by taking their rescaled Euclidean (chordal) distance to a given boundary point and then run univariate local linear regressions, and \citet{BCI14} take a similar Euclidean transformation. Other examples include \citet{BM22} that use the average of two scores, and \citet{BBRW09} and \citet{CM14} that use the minimum of multiple scores.

These approaches, however, have several drawbacks. Considering only subsets of the data leaves unexplored a rich set of treatment effects that could be identified under a valid multivariate RD design; see Section \ref{sec:setup}. Heterogeneous treatment effects of this kind can be important to answer policy questions, as exemplified by our empirical application in Section \ref{sec:emprical}. Besides, subsetting does not fully exploit signals from multiple dimensions and could suffer a loss of precision \citep{IW19}. Perhaps more precarious is the case where observations with the same distance to the cutoff are collapsed to the same point on the real line: transforming a multivariate score may incur unexpected behaviors of the resulting density of the effective univariate score, violating key assumptions needed for both identification and estimation. In particular, the density of the transformed univariate score may be zero at the new cutoff. In Section \ref{sec:zero-density}, we formally discuss the issue of zero density and provide a characterization of bivariate densities subject to this problem. We also show in simulations that such problem can make local linear regressions biased and undercover the true treatment effect, even when the sample size gets large. Nevertheless, these aforementioned issues that could arise when reducing a multivariate RD design to a univariate one have not received much attention in empirical RD research and are difficult to detect in practice without knowing the true data generating process (DGP).

What should an empirical researcher do in the presence of multiple scores, then? We consider, as alternatives to these existing approaches, two estimators that have gained popularity in estimating heterogeneous treatment effect under unconfoundedness, but have not been fully exploited in the empirical RD literature---honest regression forests of \citet{WA18} and local linear forests of \citet{FTAW20}. Like local linear regressions, these ensemble-tree methods construct a neighborhood (i.e., leaf) for a given cutoff, weighting observations by how frequently they fall into the same leaf as the cutoff across all the trees in the forest. Regression forests return the weighted average of outcomes within that neighborhood as the estimate, whereas local linear forests use these forest weights to run a weighted linear regression to capture smooth signals. Thus, regression forests and local linear forests are, respectively, analogues of local constant and local linear regressions, with forest weights replacing kernel weights. Since the identified conditional treatment effect at a given cutoff is simply the difference between two conditional means, we mimic the standard local linear regression approach by building two forests using treated and control observations, respectively, and estimate the conditional treatment effect by subtracting the control forest estimate from the treated counterpart, both evaluated at the same cutoff. These forest-based estimators are easy to implement, allow valid inference, and can accommodate any fixed number of scores that does not grow with the sample size.

To evaluate these nonparametric estimators, we design a systematic Monte Carlo study with DGPs built both from functional forms that we specify, and, following the proposal of \citet{AIMM21}, from Wasserstein Generative Adversarial Networks (WGANs) with a high degree of complexity that can closely mimic the observed data. We find that no single estimator dominates across all simulations: (i) local linear regressions perform well in univariate settings, but can undercover when multivariate scores are transformed into a univariate index---a common practice---possibly due to the ``zero-density" issue detailed in Section \ref{sec:zero-density}; (ii) the minimax-optimal estimator has good bias-variance properties when the bound on the second derivative of the conditional expectation function (CEF) is accurately estimated, but its performance is sensitive to the estimation of this nuisance parameter and its current implementation only accepts up to two scores; (iii) forest-based estimators are not designed for estimation at boundary points and can suffer from bias in finite sample---though local linear forests improve upon regression forests when the underlying signal exhibits strong linear trends---but their flexibility in modeling multivariate scores opens the door to a wide range of empirical applications in multivariate RD designs.

The rest of the paper proceeds as follows. In Section \ref{sec:setup}, we begin with the identification of conditional treatment effects in multivariate RD designs. In Section \ref{sec:estimation}, we review current estimation and inference practices using local linear regressions and the issue thereof, as well as the minimax-optimal estimator, followed by an introduction to honest regression forests and local linear forests. We present simulation results in Section \ref{sec:simulation} and an empirical application in Section \ref{sec:emprical}. 
Section \ref{sec:end} concludes. Appendix \ref{appendix:proof} collects the proof of Proposition \hyperref[prop:1]{1}, and Appendix \ref{appendix:supp} provides additional information on the Monte Carlo study. 

\section{Setup and Identification}
\label{sec:setup}
Suppose we have a sample of independent and identically distributed observations $\mathcal{S}_n :=\{(Y_i, X_i)\}_{i=1}^n$, where $Y_i \in \mathbb{R}$ is the outcome of interest and $X_i \in \mathcal{X} \subseteq \mathbb{R}^{d_X}$ is the $d_X$-dimensional running variable that determines the binary treatment status, $D_i\in\{0,1\}$, of the $i$-th unit according to some exogenous treatment assignment rule, $A:\mathcal{X} \to \{0,1\}$. If $d_X=1$, then $D_i=1$ if and only if $A(X_i)=\mathds{1}\{X_i\geq x^c\}=1$, where $x^c$ is the cutoff as in the standard sharp RD design with a univariate score. In higher dimensions, the treatment assignment rule is more complex. For example, \citet{KT15} study the effect of television advertising on election turnout, where the treatment status is determined by a geographic media market boundary. Following \citet{Z12}, let $$
\mathscr{B} := \{x\in\mathcal{X}: \forall \epsilon > 0, \ \exists x_0, x_1 \in \mathrm{Ball}(x, \epsilon)
    \text{ such that } A(x_0)=0\ \text{and}\ A(x_1)=1\} 
$$ denote the treatment boundary, i.e., a point $x\in\mathcal{X}$ lies on the treatment boundary if every $\epsilon$-ball around it contains both treated and control scores. In univariate designs, $\mathscr{B}$ is simply a singleton containing the one-dimensional cutoff. Let $\bigl(Y_i(1), Y_i(0)\bigr)$ denote the pair of potential outcomes with and without the binary treatment. We are interested in the following conditional treatment effect at any $x^c\in \mathscr{B}$:
\begin{align*}
    \tau(x^c):=\mathbb{E}[Y_i(1)-Y_i(0)|X_i=x^c].
\end{align*}
For each unit $i$, we only ever observe one of the potential outcomes. However, if we assume (i) the CEFs $\mathbb{E}\left[Y_i(1)|X_i=x^c\right]$ and $\mathbb{E}\left[Y_i(0)|X_i=x^c\right]$ are continuous at all $x^c \in \mathscr{B}$, and (ii) $X$ has a density, $f_X(\cdot)$, that is bounded away from 0 in some neighborhood around $x^c$ for all $x^c \in \mathscr{B}$, then we can identify $\tau(x^c)$ using neighboring observations around $x^c$:
\begin{align*}
    \text{For all $x^c \in \mathscr{B}$,}\quad\tau(x^c)=&\lim_{x \to x^c: ~A(x)=1}\mathbb{E}\left[Y_i(1)|X_i=x\right] - \lim_{x \to x^c:~ A(x)=0}\mathbb{E}\left[Y_i(0)|X_i=x\right]\\
    =&~\mathbb{E}\left[Y_i|X_i=x^c, D_i=1\right] - \mathbb{E}\left[Y_i|X_i=x^c, D_i=0\right],
\end{align*}
where $\tau(x^c)$ is identified as the difference between the conditional expectation of the observed outcome for the treated and control populations, respectively, both evaluated at $x^c$.\footnote{A detailed proof can be found in \citet[p.56]{Z12}.} Thus, under a valid multivariate RD design, we can identify the conditional treatment effect as a function of scores along the treatment boundary $\mathscr{B}$. Such heterogeneity may be of interest to the researcher.
%, \textcolor{red}{as we illustrate in the empirical application in Section \ref{sec:emprical}}
Alternatively, \citet{Z12} and \citet{KT15} discuss a summary measure of the average treatment effect along the boundary, i.e., $\mathbb{E}[Y_i(1)-Y_i(0)|X_i \in \mathscr{B}]$, which can be estimated given a sample of boundary points by averaging the estimated conditional treatment effects at those sample boundary points. However, as pointed out by \citet{NY23}, the Lebesgue measure of the boundary $\mathscr{B}$ is typically zero, and hence they consider the Hausdorff measure instead and propose a two-stage least squares (2SLS) estimator for 
the average treatment effect along the treatment boundary under the Hausdorff measure.
%where the weights depend on an approximate propensity score. 
In what follows, we focus on estimating conditional treatment effects at a given treatment boundary point $x^c\in\mathscr{B}$ and leave how exactly to aggregate these heterogeneous treatment effects to the empirical researcher.

\section{Estimation and Inference}
\label{sec:estimation}
In this section, we review the current estimation and inference practices using local linear regressions and the minimax-optimal estimator of \citet{IW19} in multivariate RD designs. We highlight the problem of the common practice of transforming multivariate scores to a univariate one and then running local linear regressions. We conclude this section with an introduction to the honest regression forests of \citet{WA18} and the local linear linear forests of \citet{FTAW20}, and discuss how they can be leveraged to estimate heterogeneous treatment effects in RD designs with multiple scores.

\subsection{Local Linear Regressions}
\label{sec:llr}
Given a boundary point $x^c \in \mathscr{B}$, estimation of $\tau(x^c)$ boils down to estimating two CEFs,
\vspace{-0.25cm}
\begin{align*}
    \mu^+(x^c)&:=\mathbb{E}\left[Y_i(1)|X_i=x^c\right]=\mathbb{E}\left[Y_i|X_i=x^c, D_i=1\right],\\
    \mu^-(x^c)&:=\mathbb{E}\left[Y_i(0)|X_i=x^c\right]=\mathbb{E}\left[Y_i|X_i=x^c, D_i=0\right].
\end{align*}

\vspace{-0.25cm}
\noindent Consider univariate designs first. The standard estimation approach is to run a local linear regression on each side of the cutoff using some weighting kernel $K_h(\cdot)$ with bandwidth $h$,
\vspace{-0.25cm}
\begin{align}
     \widehat{\mu}_{\texttt{llr}}^+(x^c;h) :=& \bm{e}_1^\intercal\left(\arg\min_{(\beta_0,\beta_1)^\intercal}\sum_{i=1}^n\mathds{1}\{X_i\geq x^c\}K_h(X_i-x^c)\left(Y_i-\beta_0-\beta_1(X_i-x^c)\right)^2\right),\notag\\
     =& \bm{e}_1^\intercal\bigl(X_+^\intercal W(X_+, x^c;h)X_+\bigr)^{-1}X_+^\intercal W(X_+,x^c;h)Y_+,\label{eqn:llr-weighted+}\\
     \widehat{\mu}_{\texttt{llr}}^-(x^c;h) :=& \bm{e}_1^\intercal\left(\arg\min_{(\beta_0,\beta_1)^\intercal}\sum_{i=1}^n\mathds{1}\{X_i< x^c\}K_h(X_i-x^c)\left(Y_i-\beta_0-\beta_1(X_i-x^c)\right)^2\right),\notag\\
     =& \bm{e}_1^\intercal\bigl(X_-^\intercal W(X_-,x^c;h)X_-\bigr)^{-1}X_-^\intercal W(X_-,x^c;h)Y_- ,\label{eqn:llr-weighted-}
\end{align}
where in \eqref{eqn:llr-weighted+}, the subscript “\texttt{llr}” refers to local linear regression, $\bm{e}_1:=[\,1\quad 0\,]^\intercal$ is the standard basis in $\mathbb{R}^2$ that picks out the first element of the minimizer (i.e., the fitted intercept term $\beta_0$, as the slope $\beta_1$ goes away when we evaluate the fitted regression at $x^c$); $X_+$ is the $n_1\times 2$ design matrix, where $n_1$ is the number of treated observations. The first column of $X_+$ contains all ones and the second column stacks all $\{X_i-x^c: X_i\geq x^c\}$. $W(X_+,x^c;h)$ is the $n_1\times n_1$ diagonal weighting matrix with the $i$-th diagonal element being the kernel weight $K_h(X_i-x^c)$ for $X_i\geq x^c$. $Y_+$ is the $n_1\times 1$ outcome vector that stacks the outcomes for all treated observations. Notations in \eqref{eqn:llr-weighted-} are defined analogously for the control observations. The treatment effect at $x^c$ is estimated as
\begin{align}
    \widehat{\tau}_{\texttt{llr}}(x^c;h):=\widehat{\mu}_{\texttt{llr}}^+(x^c;h)-\widehat{\mu}_{\texttt{llr}}^-(x^c;h).\label{eqn:llr-estimator}
\end{align}

Following \citet{IK12}, the choice of the bandwidth $h$ is usually obtained by minimizing the asymptotic expansion of 
\begin{align}
    \mathbb{E}\left[\bigl(\widehat{\tau}_{\texttt{llr}}(x^c;h)-\tau(x^c)\bigr)^2\right]\label{eqn:mse-opt}
\end{align}
with respect to $h$. However, \citet{CCT14} note that the MSE-optimal bandwidth obtained from optimizing \eqref{eqn:mse-opt}, denoted as $h^*$, is oversmoothing in the sense that $n(h^{{*}})^{5}$ does not converge to 0, leading to a bias term that must be corrected in order to conduct valid inference. They propose subtracting from $\widehat{\tau}_{\texttt{llr}}(x^c;h^*)$ an estimated bias term and then adjusting the standard error to account for the uncertainty in the bias estimation. This procedure, however, does not readily generalize to multivariate RD designs. Although \citet{Z12} proposes a similar MSE-optimal bandwidth selector in the bivariate case, valid inference still requires bias correction due to oversmoothing, which is technically challenging beyond one dimension and to our knowledge there is no such proposal in the existing literature. In empirical research involving multiple scores, the multivariate problem is usually reduced to one dimension so that univariate methods can apply. 

\subsubsection{A Problem with Transforming Multivariate Scores}
\label{sec:zero-density}
As mentioned earlier, common practices of reducing a multivariate RD design to a univariate one include subsetting the data and reducing the multivariate score to one dimension, and each has its own pitfalls. In particular, transforming a multivariate score in some cases incurs unexpected behaviors of the resulting density of the effective univariate running variable. In this section, we detail a situation that may arise under such transformation: the transformed univariate score may have zero density at the new cutoff, violating a key assumption for both the identification and estimation of treatment effects in RD designs.

We illustrate this problem for a transformation used by \cite{KT15}, who transform a bivariate score to a univariate one using rescaled Euclidean (chordal) distance to a given treatment boundary point and then estimate the treatment effect at the new cutoff $x^c=0$ using local linear regression with robust bias correction as in \citet{CCT14}. The following proposition gives a general characterization of bivariate densities that have a zero density at the effective univariate cutoff $x^c=0$, once the bivariate score is collapsed to a univariate variable by taking the Euclidean distance between the bivariate score and a given boundary point.

\vspace{.5cm}

\noindent
\textbf{Proposition 1.} \label{prop:1}\textit{Let $(X_1,X_2)$ be a continuous bivariate random variable with joint density $f_{(X_1,X_2)}(x_1,x_2)$ over the support $\Omega(X_1,X_2)$. Given any point $(c_1,c_2) \in \Omega(X_1,X_2)$, consider the bivariate transformation $(X_1,X_2)\mapsto(E,V)$ where $E:=\sqrt{(X_1-c_1)^2+(X_2-c_2)^2}$ and $V:=X_1-c_1$ with joint support 
\begin{align*}
    \Omega(E,V)=\left\{(e,v)\in\mathbb{R}^2: e\in\left[0,\max_{(x_1,x_2)\in\Omega(X_1,X_2)}\sqrt{(x_1-c_1)^2+(x_2-c_2)^2}\right], v \in [l(e),u(e)]\right\},
\end{align*}
where $l(e)$ and $u(e)$ are the lower and upper bound of $v$ that can depend on $e$.\footnote{To simplify notation, we assume without loss of generality that the lower and upper bounds of $(e,v)$ are attained so that the joint support of $(E,V)$ is a closed subset of $\mathbb{R}^2$; otherwise the closed brackets need to be changed to open brackets accordingly.} Then the Euclidean distance $E$ to the given point $(c_1, c_2)$ has zero density at the new cutoff $e=0$, $f_E(0)=0$, if and only if
\begin{align}
    &\biggl(\left.\int_{l(e)}^{u(e)}\frac{e}{\sqrt{e^2-v^2}}\left\{f_{(X_1,X_2)}(v+c_1,\sqrt{e^2-v^2}+c_2)\right.\right.\notag\\
    &\left.\left.\hspace{5cm}+f_{(X_1,X_2)}(v+c_1,-\sqrt{e^2-v^2}+c_2)\right\}dv\biggr)\right\rvert_{e=0}=0. \label{eqn:zero-density-cond}
\end{align}
}

\vspace{.5cm}

Intuitively, such transformation involves a loss of information: many distinct multivariate realizations can map to the same distance on the real line. In particular, for the collapsed variable to be zero, all coordinates of the original variable must simultaneously have zero distance, which is a rare event. We give the proof of Proposition \hyperref[prop:1]{1} and two examples under which \eqref{eqn:zero-density-cond} holds in Appendix \ref{appendix:proof}. This characterization depends on the population joint density of the scores and is specific to the transformation chosen by the researcher. 

\vspace{.5cm}
{
\begin{figure}[H]
    \centering
    \includegraphics[width=1\textwidth]{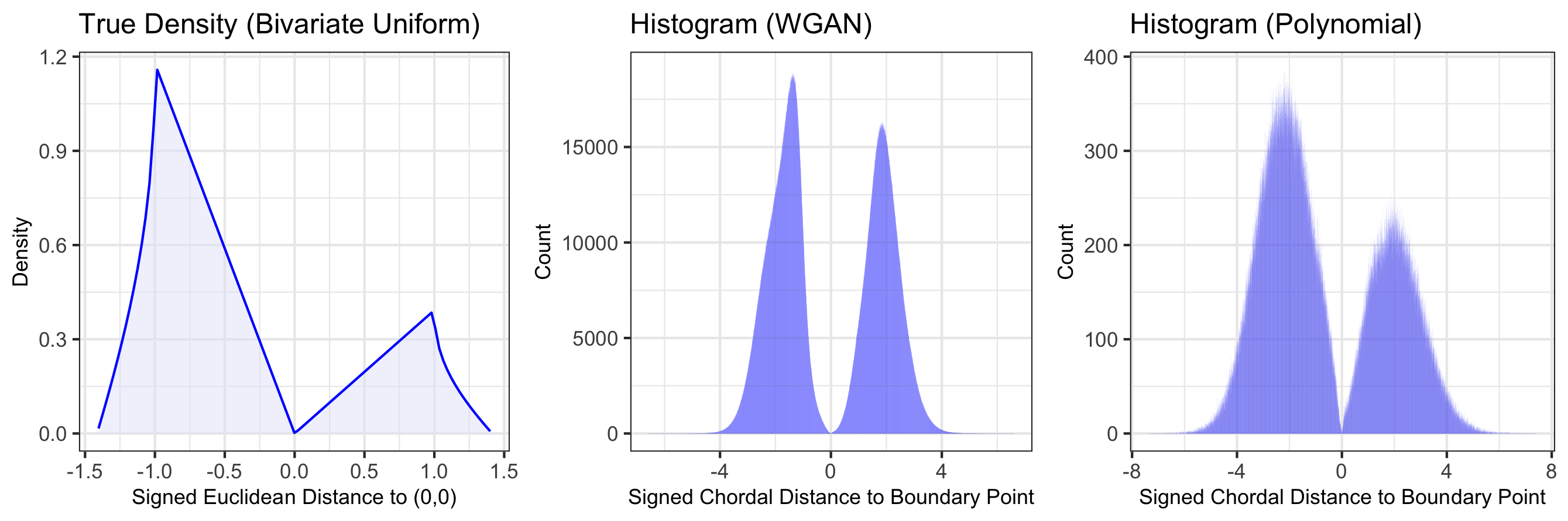}

    \vspace{-.25cm}
    \caption{True and empirical densities of transformed bivariate scores. From left to right: (i) True density of a bivariate uniform running variable on $[-1,1]^2$ transformed by signed Euclidean distance; (ii) histogram of transformed bivariate coordinates using signed chordal distance to the selected boundary point, plotted using all $40$ million observations drawn from the WGAN-generated DGP; (iii) histogram of transformed bivariate coordinates using signed chordal distance to the selected boundary point, plotted using a sample of $1$ million observations from the polynomial DGP. The number of bins for both histograms is 10,000.}
    \label{fig:density-collapsed-scores}
% {\noindent \justifying \footnotesize\textbf{Figure 1.} \par}
\end{figure}
}

To illustrate, the first panel of Figure \ref{fig:density-collapsed-scores} is the true density of a transformed bivariate score that is uniformly distributed over $[-1,1]^2$, and the point to which we calculate the Euclidean distance is $(0,0)$; see Example \hyperref[appendix:eg]{A.1} for the derivation of this density. As in \citet{KT15}, we sign the Euclidean distance so that treated (control) observations receive a positive (negative) transformed score, and the new cutoff after transformation becomes $0$, at which the density of the transformed score is exactly $0$. In practice, the researcher has access to only finite samples, and problems of this kind are difficult to detect without knowing the true DGP. In the last two panels of Figure \ref{fig:density-collapsed-scores}, we plot the raw histograms of the transformed bivariate score with respect to a given treatment boundary point using the exact same transformation employed by \citet{KT15}; each histogram uses data simulated from a different DGP built using the real data from \citet{KT15}, and thus the simulated bivariate scores do not follow any known distributions.\footnote{We provide details on how the simulation data is generated in Section \ref{sec:simulation}.} In all three panels, the empirical density of the transformed bivariate score, be it true or empirical, is zero at the effective univariate cutoff $x^c=0$. When such zero-density problem arises, not only is the conditional treatment effect at the transformed cutoff not identified because the conditional expectations are not defined, the problem is also reflected in estimation, especially for very large sample sizes, as we show in the simulations in Section \ref{sec:simulation}.

\subsection{A Minimax-Optimal Estimator}
\label{sec:minimax}
To avoid the problem of choosing a multivariate bandwidth, \citet{IW19} propose an estimator based on numerical optimization that is minimax-optimal among all estimators that are linear in the outcome variable over all problems with finite $\sigma_i^2:=Var[Y_i|X_i]$ and a second derivative bound $\mathcal{B}<\infty$ on the CEF of potential outcomes. Specifically, their estimator for $\tau(x^c)$ takes the form
\begin{align}
    \widehat{\tau}_{\texttt{mm}}(x^c;\mathcal{B}):=\sum_{i=1}^n\widehat{\gamma}_i(x^c;\mathcal{B})Y_i, \label{minimax}
\end{align}
where
\begin{align*}
    &\widehat{\gamma}(x^c;\mathcal{B})=\arg\min_{\gamma\in\mathbb{R}^n}\left\{\sum_{i=1}^n\gamma_i^2\sigma_i^2+b^2(\gamma;\mathcal{B})\right\},\\
    &b(\gamma;\mathcal{B}):=\sup_{\mu^+,\mu^-}\left\{\sum_{i=1}^n\gamma_i\mu_{A_i}(X_i)-(\mu^+(x^c)-\mu^-(x^c)): \forall\,x,\lVert\nabla^2\mu^+(x)\rVert \leq \mathcal{B}, \lVert\nabla^2\mu^-(x)\rVert\leq \mathcal{B}\right\},
\end{align*}
for $\mu_{A_i}(X_i)=\mu^+(X_i)$ if $A_i:=A(X_i)=1$ and $\mu_{A_i}(X_i)=\mu^-(X_i)$ otherwise. The norm $\lVert\cdot\rVert$ denotes the Euclidean norm when $d_X=1$ and the operator norm when $d_X>1$. In words, the estimator (\ref{minimax}) is a weighted average of the outcomes, where the weights $\widehat{\gamma}(x^c;\mathcal{B})$ are constructed by minimizing the resulting variance plus the worst-case bias over a class of CEFs with second derivatives bounded by $\mathcal{B}$. Inference follows from obtaining confidence intervals $\mathcal{I}_{\alpha}$ that achieve uniform coverage over this class of CEFs:
\begin{align*}
\mathcal{I}_{\alpha}:\liminf_{n\to\infty}\inf\left\{\mathbb{P}[\tau(x^c)\in\mathcal{I}_{\alpha}]:\forall\,x,\lVert\nabla^2\mu^+(x)\rVert \leq \mathcal{B}\text{ and } \lVert\nabla^2\mu^-(x)\rVert\leq \mathcal{B}\right\}\geq 1-\alpha.
\end{align*}
Implementing the minimax-optimal estimator (\ref{minimax}) requires estimates of $\sigma_i^2$ and $\mathcal{B}$. \citet{IW19} recommend estimating $\sigma_i^2$ by averaging the residuals from an ordinary least-squares regression of $Y_i$ on the interaction of $X_i$ and $A(X_i)$. Estimating $\mathcal{B}$, however, is problem-specific and less straightforward. \citet[p.273]{IW19} recommend estimating the CEFs globally using second-order polynomials and then multiplying the maximal estimated curvature by 2 or 4. In one of our simulations in Section \ref{sec:simulation} using data from \citet{L08}, we find that using second-order polynomials yields estimates of $\mathcal{B}$ that are too small, even after multiplying by $4$; only after multiplying the maximal estimated curvature by $30$ do we get an estimate that is close to the true (least upper) bound $\mathcal{B}$. % Which constant to use---or more generally, how exactly should one estimate $\mathcal{B}$---remains a difficult question when the researcher does not observe the ground truth in finite samples.

\subsection{Honest Regression Forests and Local Linear Forests}
\label{sec:forest}
We now introduce two additional theoretically valid estimators that can be leveraged in multivariate RD designs but have been largely overlooked in the literature---the honest regression forests of \citet{WA18} and the local linear forests of \citet{FTAW20}. Both are variants of random forests originally proposed by \citet{B01} and have received popularity in estimating heterogeneous treatment effects under unconfoundedness. 

\subsubsection{Honest Regression Forests}
\label{sec:rf}
A regression forest evaluated at some point $x \in \mathcal{X}$, denoted as $RF(x)$, is an average of $B$ trees, $\{T_b(x)\}_{b=1}^B$, each fitted using a bootstrapped sample or a random subsample of $\mathcal{S}_n:=\{(Y_i, X_i)\}_{i=1}^n$. Following \citet{ATW19}, \citet{FTAW20}, and \citet{WA18}, we focus on forests built using random subsamples, for which central limit theorems are established. Given a subsample of size $s$ from $\mathcal{S}_n$, $\{(Y_{i_k}, X_{i_k})\}_{k=1}^s$, a regression tree recursively divides the sample into mutually exclusive and collectively exhaustive convex subsets called leaves. For a point $x \in \mathcal{X}$, let $L(x)$ denote the leaf that contains $x$. A tree $T$ estimates the CEF, $\mathbb{E}[Y_i|X_i=x]$, by averaging the outcomes of observations that fall into the same leaf as $x$:
\begin{align*}
    T(x) = \frac{1}{|L(x)|}\sum_{i_k: X_{i_k} \in L(x)} Y_{i_k}.
\end{align*}
Similar to selecting an MSE-optimal bandwidth for local linear regressions, we look for partitions of the feature space that are MSE-optimal. Unlike local linear regressions, trees do not target the MSE at a particular treatment boundary point, nor do they have an analytical form for the asymptotic expansion of the overall MSE,
\begin{align*}
    \mathbb{E}\left[\bigl(Y_i-T(X_i)\bigr)^2\right],
\end{align*}
and searching for all possible partitions is computationally infeasible. Instead, standard splitting rules rely on greedy algorithms that immediately improve the in-sample fit of leaf-wide averages after each split. As such, estimates from a single tree can be highly unstable without regularization: in the extreme case, one can achieve perfect in-sample fit by partitioning the sample data so that each leaf contains only one observation. 

Regression forests stabilize predictions from individual trees by averaging many of them: for a collection of $b=1,...,B$ trees, we fit each tree $T_{b}(\cdot)$ using a random subsample of size $s$, and obtain the regression forest estimate by averaging the predictions across all $B$ trees, $\frac{1}{B}\sum_{b=1}^B T_{b}(x)$. To achieve a sizable reduction in variance, we may want to decorrelate the trees as much as possible by, for example, randomly selecting a dimension to split at each node. Following \citet{WA18}, we assume throughout that $B$ is large enough so that the Monte Carlo variability of which subsamples are drawn is negligible and adopt their definition of regression forests given the training sample $\mathcal{S}_n$ and subsample size $s$:
\begin{align*}
    RF\bigl(x;s, \mathcal{S}_n
    \bigr) := {\binom{n}{s}}^{-1}\sum_{1\leq i_1 < ... < i_s \leq n} \mathbb{E}_{\xi}\bigl[T(x;\xi, \{(Y_{i_k},X_{i_k})\}_{k=1}^s)\bigr],
\end{align*}
which averages across trees fitted using all possible draws of a size-$s$ subsample, marginalized over an auxiliary random term $\xi$ that serves as a decorrelation tool encoding, for example, the random split information. To establish asymptotic normality of regression forests, \cite{WA18} leverage a specific subsampling scheme satisfying a property called \textit{honesty} to reduce bias, which requires that the subsample used to make the splits is independent of the subsample used to estimate the within-leaf conditional means. We refer the reader to \cite{WA18} for a discussion on honesty and other regularity conditions needed to establish asymptotic normality of honest regression forests.

We now define the estimator based on honest regression forests for the conditional treatment effect at some point $x^c \in \mathscr{B}$ along the treatment boundary, $\tau(x^c)$:
\vspace{-.5cm}
\begin{align}
    \widehat{\mu}_{\texttt{rf}}^+(x^c) &:= RF\bigl(x^c; s_1, \{(Y_i,X_i) \in \mathcal{S}_n: A(X_i)=1\}\bigr),\notag\\
    \widehat{\mu}_{\texttt{rf}}^-(x^c) &:= RF\bigl(x^c; s_0, \{(Y_i,X_i) \in \mathcal{S}_n: A(X_i)=0\}\bigr),\notag\\
    \widehat{\tau}_{\texttt{rf}}(x^c)&:=\widehat{\mu}_{\texttt{rf}}^+(x^c)-\widehat{\mu}_{\texttt{rf}}^-(x^c).\label{eqn:rf-estimator}
\end{align}
Just like estimation using local linear regressions, $\widehat{\tau}_{\texttt{rf}}(x^c)$ takes the difference between a regression forest fitted using observations that fall into the treated region and a forest fitted using observations that fall into the control region, both evaluated at the same treatment boundary point $x^c\in\mathscr{B}$.

\vspace{.3cm}

\noindent\textbf{Remark 1.} 
\label{remark1} 
Although it seems that the local linear estimator (\ref{eqn:llr-estimator}) depends on one more parameter, $h$, than the regression forest estimator (\ref{eqn:rf-estimator}), we note that this is not the case. Each tree $T$ in the regression forest implicitly defines a neighborhood $L(x^c)$ for the treatment boundary point $x^c$. These neighborhoods are then aggregated, producing a weight for each observation that is proportional to the fraction of time it falls into the same leaf as $x^c$ across all trees in the regression forest. These forest weights are analogous to the weights generated by a kernel with bandwidth $h$. Thus (\ref{eqn:rf-estimator}) implicitly depends on some “pseudo-bandwidth” generated by tree partitions across the forest. Analogously, the $\gamma$ weights of the minimax-optimal estimator (\ref{minimax}) play the role of kernels and bandwidths. We can equivalently express regression forests as solutions to weighted least-squares constant regressions:
\vspace{-.5cm}
\begin{align}
    \widehat{\mu}_{\texttt{rf}}^+(x^c) &= \arg\min_{\theta}\sum_{i=1}^nD_iW_{\texttt{rf}}^+(X_i,x^c)\left(Y_i-\theta\right)^2,\notag\\
    &= \left(\mathbf{1}_{n_1}^\intercal W_{\texttt{rf}}^+(X_+,x^c)\mathbf{1}_{n_1}\right)^{-1}\mathbf{1}_{n_1}^\intercal W_{\texttt{rf}}^+(X_+,x^c)Y_+,\label{eqn:8}\\
    \widehat{\mu}_{\texttt{rf}}^-(x^c) &= \arg\min_{\theta}\sum_{i=1}^n(1-D_i)W_{\texttt{rf}}^-(X_i,x^c)\left(Y_i-\theta\right)^2,\notag\\
    &= \left(\mathbf{1}_{n_0}^\intercal W_{\texttt{rf}}^-(X_-,x^c)\mathbf{1}_{n_0}\right)^{-1}\mathbf{1}_{n_0}
    ^\intercal W_{\texttt{rf}}^-(X_-,x^c)Y_-,\label{eqn:9}
\end{align}
where the elements in (\ref{eqn:8}) and (\ref{eqn:9}) are analogous to the terms in (\ref{eqn:llr-weighted+}) and (\ref{eqn:llr-weighted-}), except $X_+$ and $X_-$ are now $n_1\times 1$ and $n_0\times 1$ vectors of ones, denoted respectively by $\mathbf{1}_{n_1}$ and $\mathbf{1}_{n_0}$,  and we replace the kernel weights with forest weights, 
\begin{align}
    W_{\texttt{rf}}^+(X_i,x^c):=\frac{1}{B}\sum_{b=1}^B\frac{\mathds{1}\{X_i\in L^+_b(x^c)\}}{|L^+_b(x^c)|}, \label{eqn:forest-weights}
\end{align}
where $L^+_b(x^c)$ denotes the leaf containing $x^c$ in the $b$-th tree fitted using the treated observations; $W_{\texttt{rf}}^-$ is defined similarly for the control observations. More generally, the least-square representation of forests in (\ref{eqn:8}) and (\ref{eqn:9}) effectively views $\widehat{\mu}_{\texttt{rf}}^+(x^c)$ and $\widehat{\mu}_{\texttt{rf}}^-(x^c)$ as solutions to $\mathbb{E}\left[(Y_i-\theta)^2|X_i=x^c, D_i=1\right]=0$ and $\mathbb{E}\left[(Y_i-\theta)^2|X_i=x^c, D_i=0\right]=0$, and solves for $\theta$ using their empirical analogs with forest weights reflective of the importance of each observation $i$ in terms of minimizing the objective function. \citet{ATW19} use forests as weighting kernels to solve parameters identified by local moments in a more general setting.

\subsubsection{Honest Local Linear Forests}
\label{sec:llf}
Though a powerful nonparametric tool and popular in applications, regression forests resemble local \textit{constant} regressions with convex weights, and hence cannot extrapolate and are known to have limited ability to capture smoothness or linear trends in the CEF, as noted by \citet{FTAW20} and the references therein, especially at points near the boundary of the support where we observe data asymmetrically from only one side of a given boundary point. To improve the performance of forests in the presence of strong linear signals, \citet{FTAW20} build on the honest regression forests of \cite{WA18} and propose a linear regression adjustment by running a ridge local linear regression with forest weights, under which we have the following estimator for $\tau(x^c)$:
\begin{align}    &\text{\small$\widehat{\mu}_{\texttt{llf}}^+(x^c):= \bm{e}_1^\intercal\left(\arg\min_{(\beta_0,\beta_1)^\intercal}\sum_{i=1}^n D_iW_{\texttt{rf}}^+(X_i,x^c)\bigl(Y_i-\beta_0-\beta_1^\intercal(X_i-x^c)\bigr)^2+\lambda\|\beta_1\|_2^2\right)$},\label{eqn:llf+}\\
&\text{\small$\widehat{\mu}_{\texttt{llf}}^-(x^c) := \bm{e}_1^\intercal\left(\arg\min_{(\beta_0,\beta_1)^\intercal}\sum_{i=1}^n (1-D_i)W_{\texttt{rf}}^-(X_i,x^c)\bigl(Y_i-\beta_0-\beta_1^\intercal(X_i-x^c)\bigr)^2+\lambda\|\beta_1\|_2^2\right)$},\label{eqn:llf-}\\
&\text{\small$\widehat{\tau}_{\texttt{llf}}(x^c):=\widehat{\mu}_{\texttt{llf}}^+(x^c)-\widehat{\mu}_{\texttt{llf}}^-(x^c)$},\label{eqn:llf-estimator}
\end{align}
where, similar to the expressions in \eqref{eqn:llr-weighted+} and \eqref{eqn:llr-weighted-} for local linear regressions, $\bm{e}_1$ is the standard basis in $\mathbb{R}^{(d_X+1)}$ with the first element being $1$ and $0$ elsewhere to pick out the fitted intercept $\beta_0$; $W_{\texttt{rf}}^+(X_i,x^c)$ and $W_{\texttt{rf}}^-(X_i,x^c)$ are the forest weights defined in \eqref{eqn:forest-weights},\footnote{To account for the fact that local linear forests will use linear regression adjustments to estimate conditional means (as opposed to using simple weighted means, as in the case of regression forests), \citet{FTAW20} also modify the standard splitting rules described in Section \ref{sec:rf} to ``ridge residual splitting;" see Section 2.1 of \citet{FTAW20}.} effectively replacing the kernel weights in \eqref{eqn:llr-weighted+} and \eqref{eqn:llr-weighted-}; $\lambda\|\beta_1\|_2^2$ is the ridge penalty term that prevents overfitting to local trends, and as explained in \citet{FTAW20}, is needed both to improve finite-sample performance and for proving convergence rates to establish a central limit theorem for local linear forests. Just like local linear regressions and regression forests, local linear forests \eqref{eqn:llf+} and \eqref{eqn:llf-} can also be expressed as solutions to weighted least-square problems; see Equation (5) of \citet[p. 507]{FTAW20}.

The assumptions needed for the asymptotic normality of local linear forests largely overlap with those required for honest regression forests to be asymptotically normal. However, with an additional assumption on the smoothness of the CEF $\mathbb{E}[Y_i|X_i=x]$---that it is differentiable with a Lipschitz continuous derivative, as opposed to the \citet{WA18} condition that only requires it to be Lipschitz continuous---\citet{FTAW20} show an improved convergence rate of local linear forests relative to the rate of regression forests, effectively exploiting the additional smoothness assumption with a linear adjustment. We refer the reader to \citet{WA18} and \cite{FTAW20} for details on the technical assumptions and discussions thereof. Pointwise confidence intervals for both honest regression forests and honest local linear forests are constructed using variance estimates from the bootstrap of little bags algorithm of \citet{SL09}, with consistency results established by the random forest delta method of \citet[Theorem 6]{ATW19}. In our Monte Carlo study, we find that local linear forests are in general less biased than regression forests and have better coverage rates whenever the simulation DGP exhibits strong linear trends.

\vspace{.5cm}

\noindent\textbf{Remark 2.} 
\label{remark2} One caveat of using local linear forests in multivariate RD designs is that the point at which we evaluate local linear forests, $x^c\in\mathscr{B}$, is necessarily a boundary point of the support of $X_i$ for both treated and control populations. On the other hand, the central limit theorem of \citet[Theorem 1]{FTAW20} only applies to points in the interior of the support, as the proof starts with a usual Taylor expansion of $\mathbb{E}[Y_i|X_i=x]$ around $x^c$.\footnote{Asymptotic normality of local linear regressions also relies on a similar Taylor expansion, but it is the CEFs of \textit{potential} outcomes that get expanded (as opposed to the CEFs of \textit{observed} outcomes of either the treated or the control population), and so in this case $x^c\in\mathscr{B}$ is an interior point of the support of $X$.} We sidestep this problem by adding an arbitrarily small ``buffer" to $x^c\in\mathscr{B}$: we pick another point $\Tilde{x}^c$ in the interior of the treated score region such that $\|\Tilde{x}^c-x^c\|=\epsilon$ for an infinitesimal $\epsilon>0$ and evaluate the treated local linear forest $\widehat{\mu}^+_{\texttt{llf}}$ at $\Tilde{x}^c$ to approximate $\widehat{\mu}^+_{\texttt{llf}}(x^c)$; we do the same for the control local linear forest $\widehat{\mu}^-_{\texttt{llf}}$. Under the assumptions in \citet[Theorem 1]{FTAW20}, both $\mu^+(x^c)$ and $\mu^-(x^c)$ are Lipschitz in $x^c$ and therefore very close to $\mu^+(\Tilde{x}^c)$ and $\mu^-(\Tilde{x}^c)$, respectively, for an infinitesimal $\epsilon$. Ideally, one should adjust the confidence interval accordingly to account for such approximation. We leave this for future research, and in our simulations and empirical application we choose $\epsilon=10^{-30}$ so that such correction is essentially negligible up to machine precision.

\section{Simulations}
\label{sec:simulation}
In this section, we conduct a systematic Monte Carlo study to evaluate the two existing estimation and inference practices in multivariate RD designs, as well as the two forest-based estimators that have not received much attention in the empirical RD literature before. All replication files can be accessed at \href{https://github.com/yqi3/GRF-RD}{\texttt{github.com/yqi3/RDForest}}.

\subsection{Data}
\label{sec:data}
We first consider a univariate design simulated using data from \citet{L08}, which is a popular choice in the RD literature to conduct simulations. \citet{L08} studies the effect of being the current incumbent party in a district on the votes obtained in the district’s next election. The running variable is univariate and measures the difference in vote share between the Democratic candidate and their strongest opponent in the current election, and the outcome variable is the Democratic vote share in the next election. This univariate design normalizes the assignment cutoff to $0$, and units with $X_i\geq 0$ are treated (i.e., winning the current election).

We then consider a bivariate design simulated using data from \citet{KT15}, who study the effect of television advertising on election turnout, where the running variable is the two-dimensional geographic coordinate of a voter's location and the outcome of interest is election turnout and takes binary values. Treatment status is determined by a geographic media market boundary. Their data also contains several other outcome variables for placebo analysis. In addition to election turnout, we include age and housing price per square foot in our simulation to cover both discrete and continuous outcome variables. \cite{KT15} consider three equally spaced points along the treatment boundary, depicted as black squares in Figure \ref{fig:KT-turnout}, to be the points at which they estimate the treatment effect. For simplicity, we pick the middle point, circled in green in Figure \ref{fig:KT-turnout}, to be the boundary point of interest in our simulation.

We consider designs up to two scores in our Monte Carlo study because the current \texttt{R} implementation of the minimax-optimal estimator $\widehat{\tau}_{\texttt{mm}}(x^c;\mathcal{B})$ in  \eqref{minimax} provided by \citet{IW19} only accepts up to two scores. In an empirical application in Section \ref{sec:emprical}, we illustrate how to use the forest-based estimators in an RD design with more than two scores to explore heterogeneous treatment effects along the treatment boundary.

\subsubsection{Generating Simulation Data}
\label{sec:buildDGP}
We consider data generating processes built both from functional forms that we specify, and, following the proposal of \citet{AIMM21}, from Wasserstein Generative Adversarial Networks (WGANs) with a high degree of complexity that can closely mimic the observed data. We first fit the observed data with a polynomial: \citet{IK12}, \citet{CCT14}, and  \citet{CCFT19} all use the \citet{L08} data to generate simulation data by fitting two $5$-th order global polynomials separately for the treated and control populations, which we also use in our simulation. For the \citet{KT15} data, we fit a $3$-rd order interacted polynomial for continuous outcomes (i.e., age and housing price) and a logistic regression using a $3$-rd order interacted polynomial inside the logit link function for the binary outcome turnout, separately for the treated and control populations. Details on these polynomial DGPs can be found in Appendix \ref{appendix:polyDGP}. 

As argued by \citet{AIMM21}, Monte Carlo studies are more credible when the data used to evaluate the performance of econometric methods does not come from researcher-chosen functional forms. Following their proposal, we also consider simulation data generated from WGAN, which trains two neural networks---one called the \textit{generator} and the other called the \textit{critic}---in an adversarial setting to minimize the Wasserstein distance between a real dataset and the artificial dataset generated from the neural nets. We refer the reader to \citet{AIMM21} for an introduction to WGAN.

Using WGAN to generate data to evaluate estimators for treatment effects, however, is task-specific, since we never directly observe the treatment effect in the data, but rather we ``identify" it under a set of identification assumptions. \citet{AIMM21} provide an example of how to generate simulation data to evaluate estimators of treatment effects under unconfoundedness, where they train two conditional WGANs to learn $X_i|D_i$ and $Y_i|(X_i, D_i)$ from the real sample. They then draw an artificial sample of treatment status and covariates from the WGAN-generated distribution of $X_i|D_i$, and for each $(X_i, D_i)$ drawn, an outcome is drawn from the WGAN-generated distribution of $Y_i|(X_i, D_i)$. Under unconfoundedness, $Y_i|(X_i, D_i=d)$ identifies $Y_i(d)|X_i$ and hence for each $Y_i$ drawn, its counterfactual outcome is generated by mutating the corresponding treatment status $D_i$ to $(1-D_i)$ but keeping the same $X_i$. In other words, the counterfactual outcome is drawn from the WGAN-generated distribution of $Y_i|(X_i, 1-D_i)$. The true average treatment effect is simulated by averaging the difference between the pair of generated outcome and its counterfactual over a large WGAN-generated sample.

\begin{figure}[H]
\centering
\includegraphics[width=1\textwidth]{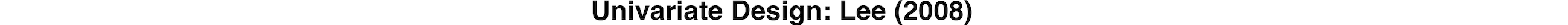}
    \includegraphics[width=0.31\textwidth]{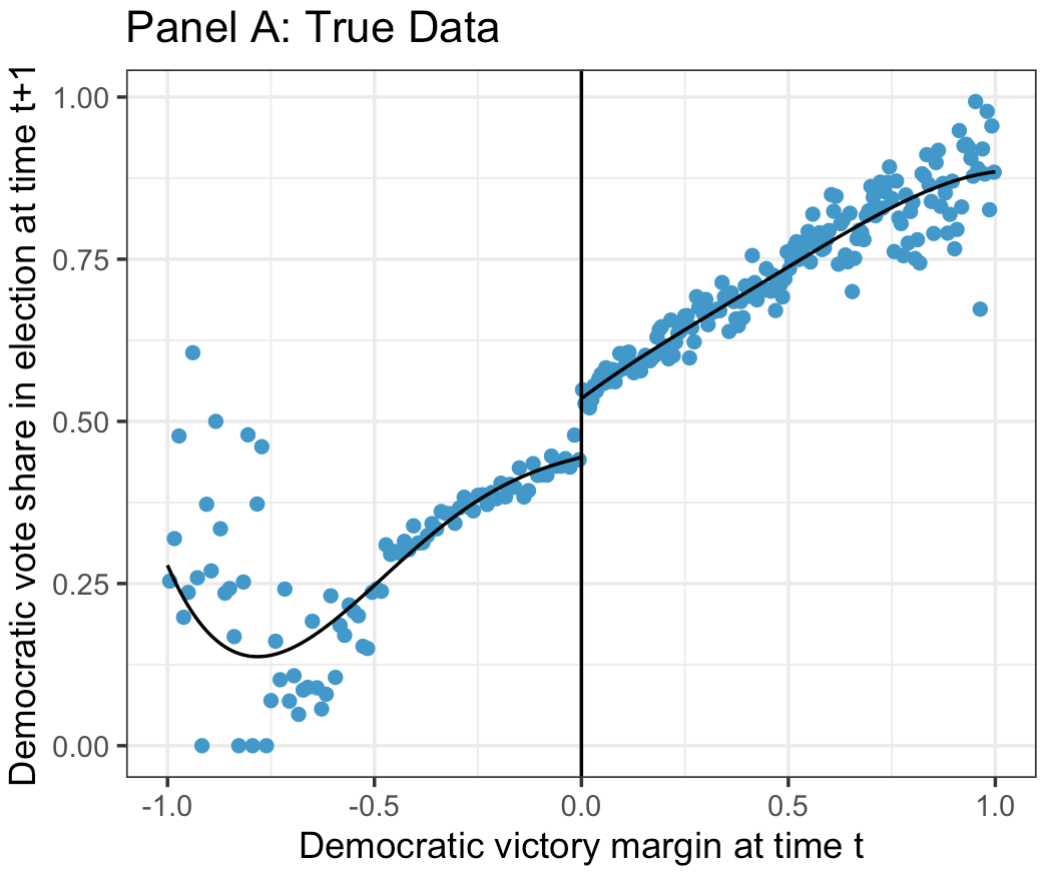}
    \includegraphics[width=0.31\textwidth]{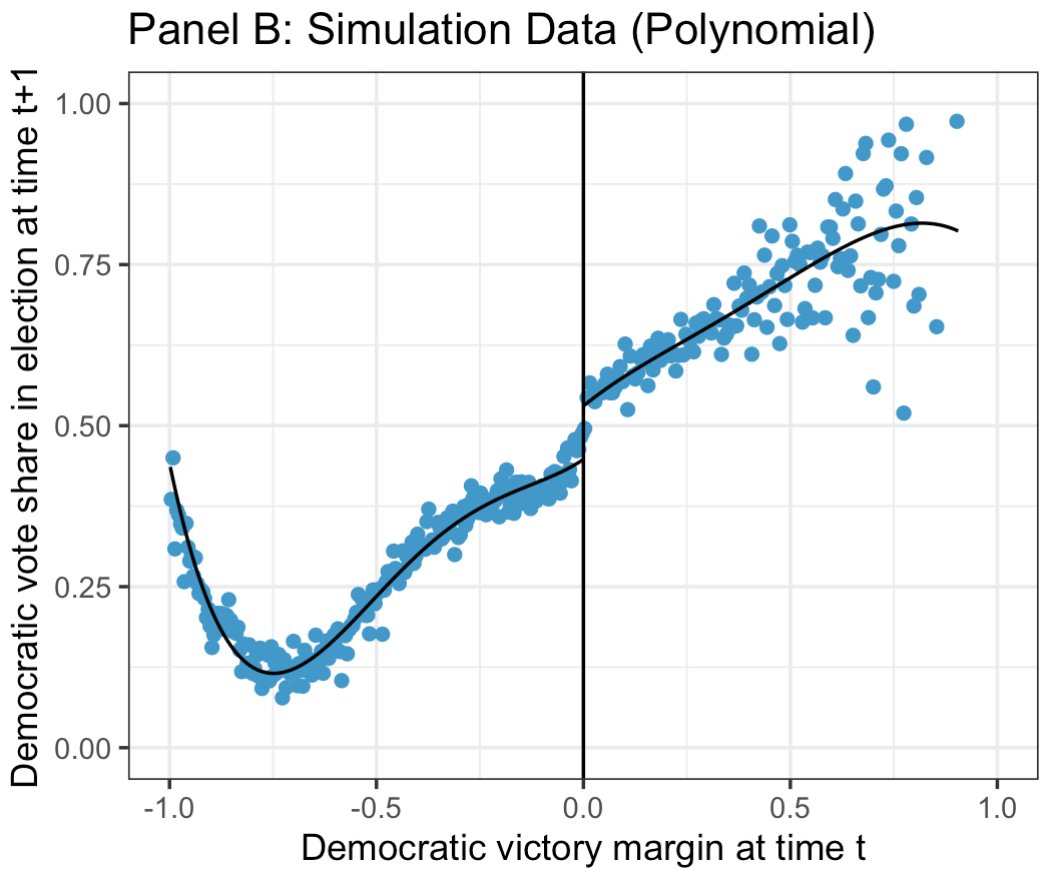}
    \includegraphics[width=0.31\textwidth]{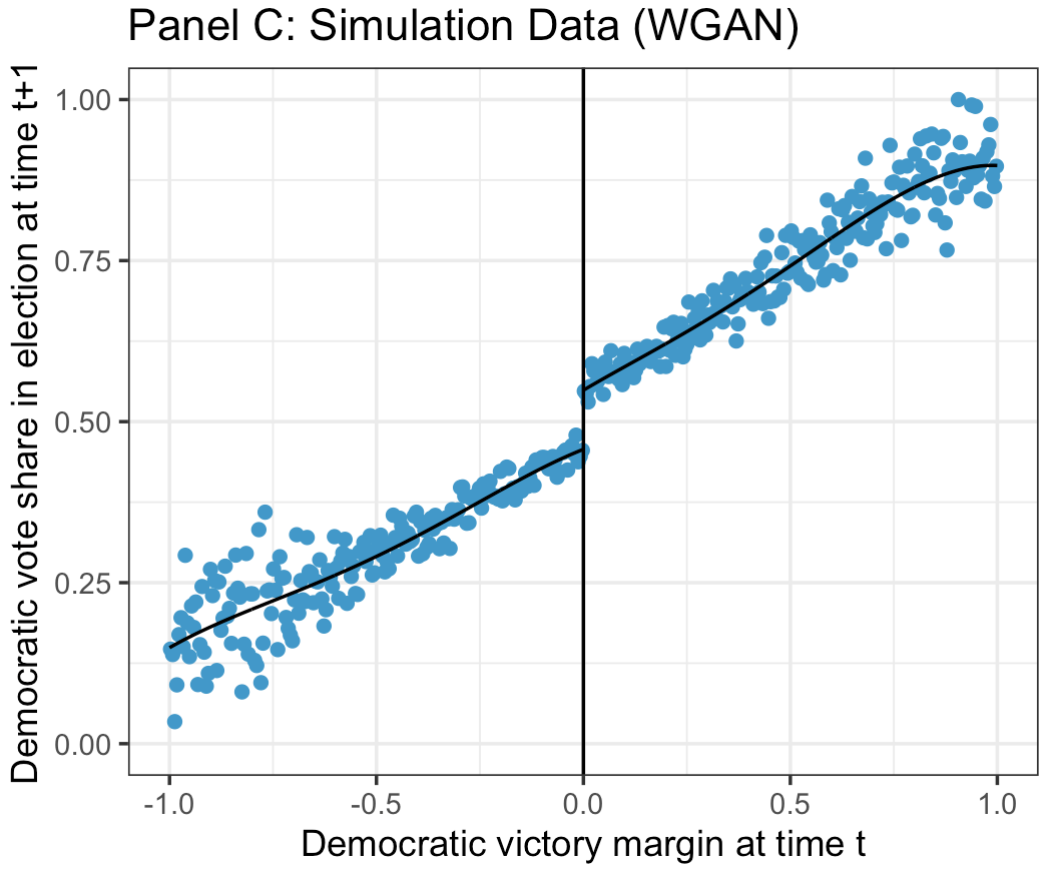}
    \caption{\scriptsize \citet{L08}: Democratic victory margin in the current election against the vote share in the next election. Panel A shows the observed data. Panel B plots the model data fitted using a 5-th order polynomial with different coefficients for the treated and control observations. Panel C plots the WGAN data. The cutoff is normalized to 0. The figure is generated using the \texttt{rdrobust} package with default smoothing \citep{CCT14}.}
    \label{fig:Lee}
\end{figure}

\begin{figure}[H]
    \centering
    \includegraphics[width=0.95\textwidth]{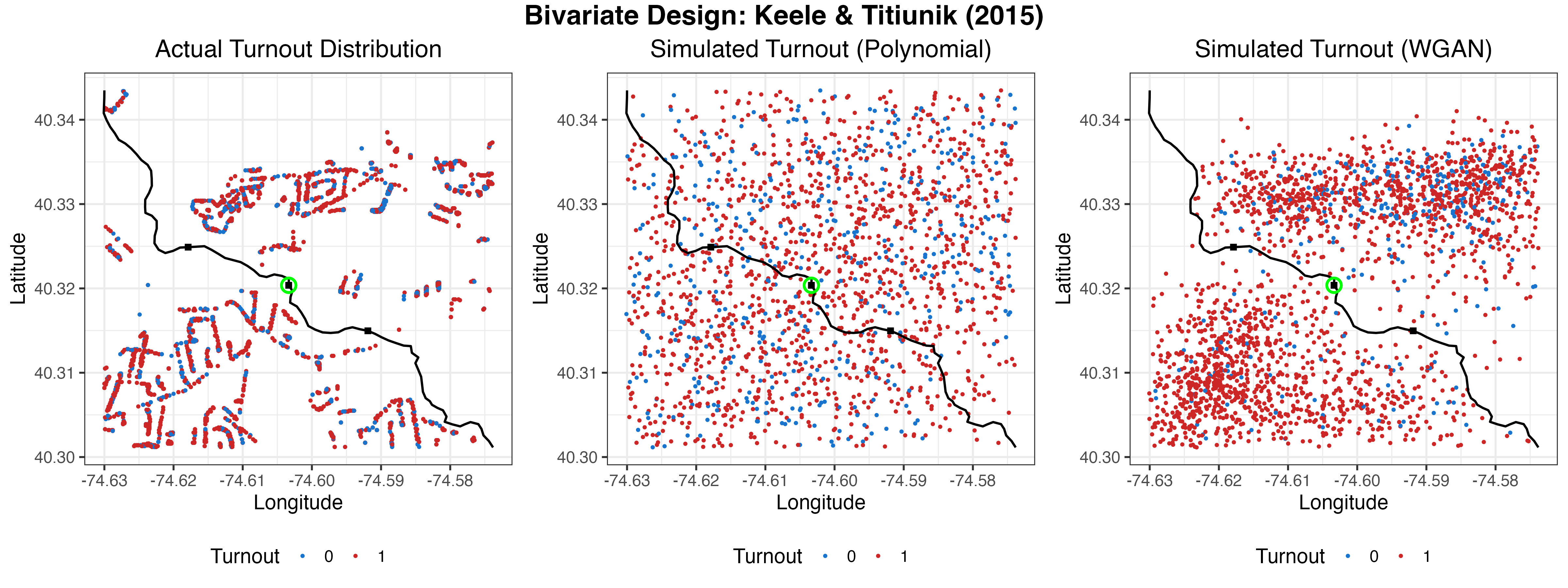}
    \vspace{-0.25cm}
    \caption{\scriptsize \citet{KT15}: Geographic coordinates and election turnout. The leftmost panel shows the real data, and the middle and rightmost panels visualize respectively the simulated polynomial and WGAN DGPs.  Coordinates below the black curve (treatment boundary) are treated, and the green circle shows the boundary point at which we estimate the treatment effect. To visualize the polynomial DGPs, 2,500 coordinates are sampled with replacement from the original sample plus a noise term drawn from $\mathcal{N}(0,0.01^2)$.    
    %Each realization of turnout follows a Bernoulli distribution, where the parameter is the likelihood given by a logit model evaluated at the corresponding coordinates. 
    }
    \label{fig:KT-turnout}
\end{figure}

Our setting, with a different design and identification assumption for the treatment effect of interest, requires a different way of using WGAN to generate suitable simulation data. Specifically, the treatment effects are only identified for  $X_i\in\mathscr{B}$, so we need to simulate the treatment effect at each treatment boundary point of interest, instead of simply taking averages as in the unconfoundedness setting. To do so, we first train two conditional WGANs to learn $X_i|D_i$ and $Y_i|(X_i, D_i)$ from the real sample, as in \citet{AIMM21}. But for a given treatment boundary point $x^c\in\mathscr{B}$, we draw a separate artificial sample to simulate the true treatment effect at $x^c$: after training the WGAN to learn the distribution of $Y_i|(X_i, D_i)$, we generate 100,000 outcomes from $Y_i|(X_i=x^c, D_i=1)$ and another 100,000 outcomes from $Y_i|(X_i=x^c, D_i=0)$, and take the average of their differences as the true treatment effect at $x^c$.\footnote{We note that WGANs do not directly yield closed-form estimates of the fitted distribution, but rather generate a new, artificial sample of arbitrary size that closely mimics the real sample. This is why we need to \textit{simulate} the true treatment effect at a given treatment boundary point $x^c\in\mathscr{B}$, unlike in the polynomial case where we can directly calculate the true treatment effects from the coefficients of the fitted polynomials evaluated at $x^c$.} Figures \ref{fig:Lee} and \ref{fig:KT-turnout} visualize the actual data and the simulated data from the polynomial specification and WGAN. In Appendix \ref{appendix:supp}, we include additional descriptions of the WGAN-generated DGPs in Figures \ref{fig:KT-price&age}--\ref{fig:surface main}, and we report in Table \ref{tbl:wd} the Wasserstein distance between the real and WGAN data. We use the default options available from the \texttt{wgan} package of \citet{AIMM21} to train the WGANs; see Appendix \ref{appendix:wganDGP} for details. Following Section 5 of \citet{AIMM21}, we conduct a thorough robustness check to assess the variability of the Monte Carlo results with respect to sampling variation and WGAN tuning parameters, with robustness results reported in Appendix \ref{appendix:wganRobust}.

\subsection{Monte Carlo Results}
\label{sec:MCresults}
We report the simulation results in Table \ref{tbl:mc-results}. All simulations are implemented in \texttt{R}, where we use the \texttt{rdrobust} package of \cite{CCT14} for local linear regressions, the \texttt{optrdd} package of \cite{IW19} for the minimax-optimal estimator, and the \texttt{grf} package of \cite{ATW19} for the two forest estimators---all with default options without tuning, except we specify pilot values for a few hyperparameters of the two forest estimators, for reasons explained in Appendix \ref{appendix:forest-param}.

\begin{landscape}
\begin{table}
\centering
\resizebox{1\columnwidth}{!}{
\begin{tabular}
{@{\extracolsep{3pt}}l@{}c@{}@{}c@{}@{}c@{}@{}c@{}@{}c@{}@{}c@{}@{}c@{}@{}c@{}@{}c@{}@{}c@{}@{}c@{}@{}c@{}@{}c@{}@{}c@{}@{}c@{}@{}c@{}@{}c@{}@{}c@{}@{}c@{}@{}c@{}@{}c@{}@{}c@{}@{}c@{}} 
 & \multicolumn{5}{c}{Local Linear Regression} & \multicolumn{1}{c}{}
 & \multicolumn{5}{c}{Minimax-Optimal} & \multicolumn{1}{c}{} & \multicolumn{5}{c}{Honest Regression Forest} & \multicolumn{1}{c}{} & \multicolumn{5}{c}{Honest Local Linear Forest}\\
\cline{2-6} \cline{8-12} \cline{14-18} \cline{20-24}
\\[-2ex] & (1) & (2) & (3) & (4) & (5) & & (6) & (7) & (8) & (9) & (10) & & (11) & (12) & (13) & (14) & (15) & & (16) & (17) & (18) & (19) & (20)
\vspace{.2cm}
\\
\multicolumn{1}{r}{Sample size $n=$} & $1,000$ & $5,000$ & $10,000$ & $50,000$ & $100,000$ & & $1,000$ & $5,000$ & $10,000$ & $50,000$ & $100,000$ & & $1,000$ & $5,000$ & $10,000$ & $50,000$ & $100,000$ & & $1,000$ & $5,000$ & $10,000$ & $50,000$ & $100,000$
\\[-.1ex]\hline
\\[-.5ex]
\multicolumn{24}{c}{\large \textbf{Univariate Design: Polynomial DGP}}
\\\hline
\multicolumn{24}{l}{\cellcolor{blue!20}\textbf{Panel A: Vote Share (truth $\boldsymbol{=0.0400}$)}}
\vspace{.1cm}\\
\multicolumn{1}{l}{Bias} & 0.0161 & 0.0116 & 0.0078 & 0.0026 & 0.0011 &\vline & 0.0378 & 0.0257 & 0.0211 & 0.0134 & 0.0102 &\vline & 0.0344 & 0.0072 & 0.0043 & 0.0006 & 0.0019 &\vline & 0.0180 & 0.0051 & 0.0036 & 0.0008 & 0.0019\\
\multicolumn{1}{l}{Variance} & 0.0021 & 0.0005 & 0.0003 & 0.0001 & 0.0000 &\vline & 0.0008 & 0.0002 & 0.0001 & 0.0000 & 0.0000 &\vline & 0.0014 & 0.0013 & 0.0013 & 0.0015 & 0.0013 &\vline & 0.0030 & 0.0020 & 0.0019 & 0.0017 & 0.0014\\
\multicolumn{1}{l}{95\% coverage} & 91.1\%	& 86.8\% & 89.1\% & 91.2\% & 92.9\% &\vline & 80.5\% & 68.7\%	& 64.8\% & 49.8\% & 51.6\% &\vline & 85.1\% & 94.9\% & 95.6\% & 94.4\% & 94.4\% &\vline & 88.9\% & 92.0\% & 94.0\% & 93.1\% & 93.8\%\\
\hdashline
\\[-.5ex]
\multicolumn{24}{c}{\large \textbf{Univariate Design: WGAN DGP}}
\\\hline
\multicolumn{24}{l}{\cellcolor{blue!20}\textbf{Panel B: Vote Share (truth $\boldsymbol{=0.0992}$)}}
\vspace{.1cm}\\
\multicolumn{1}{l}{Bias} & 0.0013 & 0.0013 & 0.0013 & 0.0014 & 0.0016 &\vline & -0.0031 & -0.0045 & -0.0042 & -0.0033 & -0.0025 &\vline & 0.0102 & 0.0023 & 0.0017 & 0.0000 & 0.0028 &\vline & 0.0055 & 0.0010 & 0.0015 & -0.0001 & 0.0027 \\
\multicolumn{1}{l}{Variance} & 0.0007 & 0.0002 & 0.0001 & 0.0000 & 0.0000 &\vline & 0.0002 & 0.0000 & 0.0000 & 0.0000 & 0.0000 &\vline & 0.0009 & 0.0008 & 0.0010 & 0.0008 & 0.0008 &\vline & 0.0015 & 0.0012 & 0.0012 & 0.0008 & 0.0008\\
\multicolumn{1}{l}{95\% coverage} & 94.1\% & 93.5\% & 95.5\% & 93.9\% & 91.6\% &\vline & 95.7\% & 93.2\% & 92.5\% & 76.6\% & 72.7\% &\vline & 95.7\% & 96.1\% & 94.2\% & 96.6\% & 95.2\% &\vline & 93.1\% & 94.1\% & 92.9\% & 96.2\% & 95.2\%
\\
\hdashline
\\[-.5ex]
\multicolumn{24}{c}{\large \textbf{Bivariate Design: Polynomial DGP}}
\\\hline
\multicolumn{24}{l}{\cellcolor{blue!20}\textbf{Panel C: Turnout Probability (truth $\boldsymbol{=-0.0376}$)}}
\vspace{.1cm}\\
\multicolumn{1}{l}{Bias} & -0.0118 & -0.0201 & -0.0120 & -0.0185 & -0.0161 &\vline & -0.0049 & -0.0043 & 0.0013 & -0.0011 & -0.0009 &\vline & 0.0220 & 0.0134 & 0.0117 & 0.0078 & 0.0011 &\vline & 0.0169 & 0.0127 & 0.0106 & 0.0079 & 0.0012\\
\multicolumn{1}{l}{Variance} & 0.0786 & 0.0127 & 0.0051 & 0.0010 & 0.0005 &\vline & 0.0189 & 0.0060 & 0.0034 & 0.0011 & 0.0007 &\vline & 0.0070 & 0.0063 & 0.0060 & 0.0061 & 0.0060 &\vline & 0.0084 & 0.0064 & 0.0060 & 0.0059 & 0.0058\\
\multicolumn{1}{l}{95\% coverage} & 93.4\% & 94.8\% & 95.3\% & 93.1\% & 93.3\% &\vline & 97.8\% & 97.3\% & 98.1\% & 98.3\% & 97.1\% &\vline & 95.3\% & 94.9\% & 95.2\% & 94.7\% & 93.6\% &\vline & 94.1\% & 94.9\% & 94.8\% & 94.0\% & 93.7\%\\
\\[-2ex] \hline 
\multicolumn{24}{l}{\cellcolor{blue!20}\textbf{Panel D: Housing Price (truth $\boldsymbol{=1.4878}$)}}
\vspace{.1cm}\\
\multicolumn{1}{l}{Bias} & -3.8353 & -4.0482 & -3.1878 & -1.9335 & -1.2293 &\vline	& 0.8840 & -0.2467 & -0.5726 & -0.2619 & -0.1799 &\vline & 6.7977 & 3.9785 & 3.4101 & 1.9859 & 2.1574 &\vline & 1.2121 & 0.8221 & 0.8680 & 0.0998 & 0.6290\\
\multicolumn{1}{l}{Variance} & 528.8758 & 85.8068 & 37.4150 & 9.2601 & 5.7001 &\vline & 315.3792 & 95.7504 & 59.4647 & 15.2623 & 8.8313 &\vline & 49.9737 & 37.6548 & 39.3580 & 34.7354 & 37.3014 &\vline & 87.1682 & 59.4337 & 57.0010 & 46.7899 & 48.0771\\
\multicolumn{1}{l}{95\% coverage} & 94.8\% & 94.7\% & 95.5\% & 95.7\% & 95.0\% &\vline & 97.2\% & 97.8\% & 97.4\% & 97.6\% & 97.0\% &\vline & 82.4\% & 92.4\% & 91.5\% & 94.1\% & 91.9\% &\vline & 93.8\% & 94.6\% & 95.8\% & 93.2\% & 93.6\%\\
\\[-2ex]\hline 
\multicolumn{24}{l}{\cellcolor{blue!20}\textbf{Panel E: Age (truth $\boldsymbol{=1.0880}$)}}
\vspace{.1cm}\\
\multicolumn{1}{l}{Bias} & -0.0178 & -0.1874 & -0.1966 & -0.1698 & -0.1408 &\vline	& -0.2898 & 0.0052 & 0.0626 & 0.0390 & -0.0502 &\vline & -0.2220 & -0.2153 & -0.2951 & 0.0170 & -0.2594 &\vline & 0.2984 & -0.0097 & -0.1705 & 0.0631 & -0.2111\\
\multicolumn{1}{l}{Variance} & 110.6967 & 15.8490 & 7.9903 & 1.2539 & 0.6609 &\vline & 36.1169 & 9.6130 & 6.1630 & 2.1915 & 1.2345 &\vline & 9.2555 & 9.1571 & 8.4339 & 7.8535 & 7.5993 &\vline & 13.8381 & 11.5593 & 9.9945 & 8.5617 & 8.2755\\
\multicolumn{1}{l}{95\% coverage} & 93.3\% & 94.9\% & 94.2\% & 97.1\% & 95.9\% &\vline & 97.6\% & 97.6\% & 97.7\% & 97.4\% & 96.8\% &\vline & 95.7\% & 93.9\% & 94.5\% & 94.2\% & 94.7\% &\vline & 94.6\% & 93.5\% & 93.6\% & 93.7\% & 94.9\%\\
\hdashline
\\[-.5ex]
\multicolumn{24}{c}{\large \textbf{Bivariate Design: WGAN DGP}}
\\\hline
\multicolumn{24}{l}{\cellcolor{blue!20}\textbf{Panel F: Turnout Probability (truth $\boldsymbol{=0.0356}$)}}
\vspace{.1cm}\\
\multicolumn{1}{l}{Bias} & 0.0180 & 0.0294 & 0.0257 & 0.0259 & 0.0252 &\vline & -0.0061 & -0.0063 & -0.0022 & -0.0019 & -0.0020 &\vline & 0.0490 & 0.0309 & 0.0242 & 0.0211 & 0.0177 &\vline & 0.0426 & 0.0280 & 0.0217 & 0.0197 & 0.0169\\
\multicolumn{1}{l}{Variance} & 0.2461 & 0.0240 & 0.0094 & 0.0015 & 0.0008 &\vline & 0.0305 & 0.0122 & 0.0075 & 0.0028 & 0.0015 &\vline & 0.0054 & 0.0046 & 0.0043 & 0.0039 & 0.0040 &\vline & 0.0069 & 0.0051 & 0.0045 & 0.0036 & 0.0037\\
\multicolumn{1}{l}{95\% coverage} & 94.8\% & 93.6\% & 93.8\% & 90.1\% & 85.5\% &\vline & 98.5\% & 97.9\% & 98.4\% & 97.9\% & 99.1\% &\vline & 90.7\% & 93.8\% & 93.8\% & 94.9\% & 93.9\% &\vline & 86.8\% & 90.6\% & 93.1\% & 94.4\% & 93.0\%\\
\\[-2ex] \hline 
\multicolumn{24}{l}{\cellcolor{blue!20}\textbf{Panel G: Housing Price (truth $\boldsymbol{= 4.0414}$)}}
\vspace{.1cm}\\
\multicolumn{1}{l}{Bias} & 2.7574 & 4.8878 & 5.6543 & 4.5000 & 3.8748 &\vline & -0.2310 & -0.5278 & -0.0903 & -0.1227 & -0.0275 &\vline & 8.7037 & 5.7927 & 4.9030 & 2.4293 & 2.1113 &\vline & 6.3921 & 4.7853 & 4.2545 & 2.2442 & 1.9833\\
\multicolumn{1}{l}{Variance} & 3806.2308 & 232.4295 & 80.8610 & 10.4512 & 5.0567 &\vline & 423.0745 & 180.9454 & 109.4033 & 29.0582 & 14.4899 &\vline & 37.0680 & 34.4074 & 31.7393 & 27.7547 & 25.9484 &\vline & 72.5886 & 50.0642 & 44.7507 & 30.8009 & 27.3207\\
\multicolumn{1}{l}{95\% coverage} & 94.0\% & 93.4\% & 93.0\% & 88.0\% & 84.0\% &\vline & 98.1\% & 97.4\% & 97.2\% & 97.9\% & 98.7\% &\vline & 72.2\% & 81.5\% & 85.3\% & 91.9\% & 92.2\% &\vline & 80.6\% & 84.3\% & 85.5\% & 91.0\% & 93.4\%\\
\\[-2ex]\hline 
\multicolumn{24}{l}{\cellcolor{blue!20}\textbf{Panel H: Age (truth $\boldsymbol{=4.4519}$)}}
\vspace{.1cm}\\
\multicolumn{1}{l}{Bias} & -2.2493 & -1.2546 & -1.1629 & -1.2299 & -1.2656	&\vline & -0.2949 & 0.0769 & 0.0839 & -0.0428 & -0.0237 &\vline & -0.6848 & -0.1871 & -0.1209 & -0.0749 & -0.2052 &\vline & -0.7206 & -0.1786 & -0.0985 & -0.0659 & -0.2074\\
\multicolumn{1}{l}{Variance} & 2200.4077 & 58.0355 & 21.9401 & 2.8652 & 1.4544	&\vline & 68.6196 & 23.8051 & 14.7071 & 4.4742 & 2.4230 &\vline & 8.2230 & 7.2583 & 6.4681 & 6.4340 & 5.8234 &\vline & 13.4557 & 8.5387	& 7.0771 & 6.5556 & 5.9329\\
\multicolumn{1}{l}{95\% coverage} & 92.1\% & 94.4\% & 94.5\% & 91.7\% & 85.1\% &\vline & 98.1\% & 98.3\% & 98.8\% & 98.4\% & 98.3\% &\vline & 96.0\% & 95.0\% & 95.3\% & 94.3\% & 94.6\% &\vline & 93.8\% & 93.8\% & 94.9\% & 94.9\% & 95.0\%\\
\\[-2ex]\hline
\end{tabular}
}

\vspace{-.25cm}
\caption{\footnotesize Monte Carlo simulation results. For each method, we vary the sample size $n\in\{1000, 5000, 10000, 50000, 100000\}$ and report the bias, variance, and $95\%$ coverage rate across 1,000 Monte Carlo replications. Each panel corresponds to a different outcome variable in Section \ref{sec:data}, with the true treatment effect in parentheses either calculated exactly (from the polynomial fit) or simulated as described in Section \ref{sec:buildDGP} (for the WGAN DGPs).}
\label{tbl:mc-results}
\end{table}
\end{landscape} 

Local linear regressions perform well in univariate settings, although they slightly undercover in the polynomial DGP (Panel A, Columns 1-5 of Table \ref{tbl:mc-results}).\footnote{This undercoverage is consistent with the simulation result ($90.6\%$ coverage) of \cite{CCFT19}, where they use the same polynomial DGP but with $n=1000$ for a total of $5000$ replications; see Table SA-1, Row 2 and Column 3 in their supplemental appendix. We have done our simulation using their \texttt{STATA} code and obtained similar results to those reported in Table \ref{tbl:mc-results}.} However, they are generally noisier in bivariate settings, where we follow \cite{KT15} to first collapse the bivariate score using rescaled Euclidean (chordal) distance and then run \texttt{rdrobust}, especially in small samples. Although the variance of local linear regressions shrinks as the sample size gets large, their bias does not shrink fast enough relative to the diminishing variance in those bivariate WGAN DGPs (Panels F-H, Columns 1-5 of Table \ref{tbl:mc-results}), and therefore the coverage rate drops as the sample size increases. This problem arises possibly due to the zero-density issue of the collapsed multivariate score at the new univariate cutoff, as we point out in Section \ref{sec:zero-density} and empirically illustrate in the last two histograms of Figure \ref{fig:density-collapsed-scores}, cautioning against such univariate transformations.

The minimax-optimal estimator of \cite{IW19} has a relatively low bias but tends to overcover, which is expected since it is designed for minimizing the \textit{worst-case} MSE and inference aims at achieving \textit{uniform} coverage over a class of CEFs explained in Section \ref{sec:minimax}. Nevertheless, its performance is contingent on how precisely the second derivative bound $\mathcal{B}$ is estimated: for the two artificial datasets built using the \cite{L08} data (Panels A-B of Table \ref{tbl:mc-results}), the minimax-optimal estimator undercovers severely, especially for the polynomial DGP in Panel A. There, we follow the recommendation of \citet{IW19}
to estimate the CEFs globally using a second-order polynomial and multiply the maximal estimated curvature by $4$. But this turns out to yield an estimate of $\mathcal{B}$ that is too small; Table \ref{tbl:lee-B} in Appendix \ref{appendix:sec-deriv-bound} shows how the simulation results change with the constant multiplied by $\mathcal{B}$, and it is only after multiplying the maximal estimated curvature by $30$ do we get an estimate that is close to the true (least upper) bound $\mathcal{B}$.

Of course, one could argue that estimating $\mathcal{B}$ by a global second-order polynomial is a bad choice to start with. For the bivariate design, we follow their working paper \citep[p.21]{IW19w} and estimate $\mathcal{B}$ using the $95$-th percentile (over all sample points) of the operator norm curvature via a cross-validated ridge regression with interacted $7$-th order natural splines, which yields good coverage (Panels C-H, Columns 6-10 of Table \ref{tbl:mc-results}). However, without knowledge about the true DGP, choosing a “good” model is not straightforward, and estimating second derivatives is more difficult when the dimension of the running variable increases. In addition, the current implementation of \texttt{optrdd} accepts up to two scores only, limiting its application more broadly.

Honest regression forests and local linear forests have comparable performance to local linear regressions in the univariate design (Panels A-B, Columns 11-20 of Table \ref{tbl:mc-results}), though their variances are slightly higher. On the other hand, regression forests are generally more biased than local linear forests, especially in small samples, which is a known boundary problem for regression forests. In addition, the variance of regression forests tends to be smaller than that of local linear forests, exacerbating the undercoverage problem. As we discussed in Section \ref{sec:llf}, local linear forests are better at capturing smooth signals via a linear correction, and its advantage over regression forests is especially pronounced when the underlying DGP exhibits strong linear trends near the boundary point at which we estimate the treatment effect. In Appendix \ref{appendix:wganRobust}, we examine the robustness of the results in Table \ref{tbl:mc-results} to the number of training epochs, which is a tuning parameter of WGAN, and increasing the number of training epochs leads to stronger linear trends near the boundary point (see Figure \ref{fig:surface main}); in these cases, the performance of regression forests deteriorates, whereas local linear regressions have robust performance. With that said, these two forest-based methods are theoretically valid alternatives to existing methods for multivariate RD designs and can flexibly model any given number of scores, and empirical researchers may leverage and tailor them accordingly to fit different empirical settings.

\section{Empirical Application}
\label{sec:emprical}
We revisit the empirical analysis in \citet{NY23}, who study the effect of hospital relief funding during COVID-19. 

\subsection{Data}
The pandemic inflicted substantial financial pressure on U.S. hospitals, compounding revenue losses with increasing operational expenses. To alleviate this, the U.S. government enacted the Coronavirus Aid, Relief, and Economic Security (CARES) Act in March 2020 that provides emergency financial support to healthcare providers. A \$10 billion portion of this funding was allocated to ``safety net" hospitals, as determined by a set of eligibility criteria that target hospitals serving vulnerable populations. Hospitals were eligible for safety net funding if they met all of the following three conditions:\footnote{Source: \href{https://www.hrsa.gov/sites/default/files/hrsa/coronavirus/provider-relief-fund-faq-complete.pdf}{Provider Relief Fund Frequently Asked Questions} from the Health Resources and Services Administration, U.S. Department of Health and Human Services.}
\begin{enumerate}
    \item Medicare Disproportionate Patient Percentage (DPP) of 20.2\% or higher.\footnote{For a more detailed definition, see 
    \href{https://www.cms.gov/medicare/payment/prospective-payment-systems/acute-inpatient-pps/disproportionate-share-hospital-dsh}{Centers for Medicare \& Medicaid Services}.}
    
    \item Annual Uncompensated Care (UCC) exceeding \$25{,}000 per bed.\footnote{For a more detailed definition, see \href{https://www.aha.org/fact-sheets/2020-01-06-fact-sheet-uncompensated-hospital-care-cost}{Fact Sheet: Uncompensated Hospital Care Cost}.}
    
    \item Profit margin (defined as net income divided by the sum of net patient revenue and total other income) of 3.0\% or less.
\end{enumerate}
To determine safety net funding eligibility, we use data from the Healthcare Cost Report Information System (HCRIS) for the 2019 financial year \citep{TL-303}. The 2019 measurements are the closest available approximation of the values hospitals likely reported when they applied for funding in 2020.\footnote{Although using the 2020 data may seem more temporally aligned with the application timeline, it is unclear whether these records were actually used by the hospitals at the time of application. Since the coverage of the 2019 data was low at the time of \cite{NY23}'s study, they use the 2018 HCRIS data instead. We thank them for their suggestion to use the updated data and help with data preparation.} For the outcome variable, following \cite{NY23}, we use the number of confirmed or suspected COVID patients hospitalized in each hospital during the week spanning August 2 to August 8, 2020, drawn from the most recent (June 2024) version of the COVID Reported Patient Impact and Hospital Capacity by Facility dataset \citep{HHS2024}. We note that, due to changes in the reporting format, our weekly outcome is based on a slightly different weakly reporting window from that of  \cite{NY23}, which starts from July 31 to August 6, 2020 using an earlier version of the outcome dataset.

\begin{figure}[H]
\centering
   \includegraphics[width=0.7\textwidth]{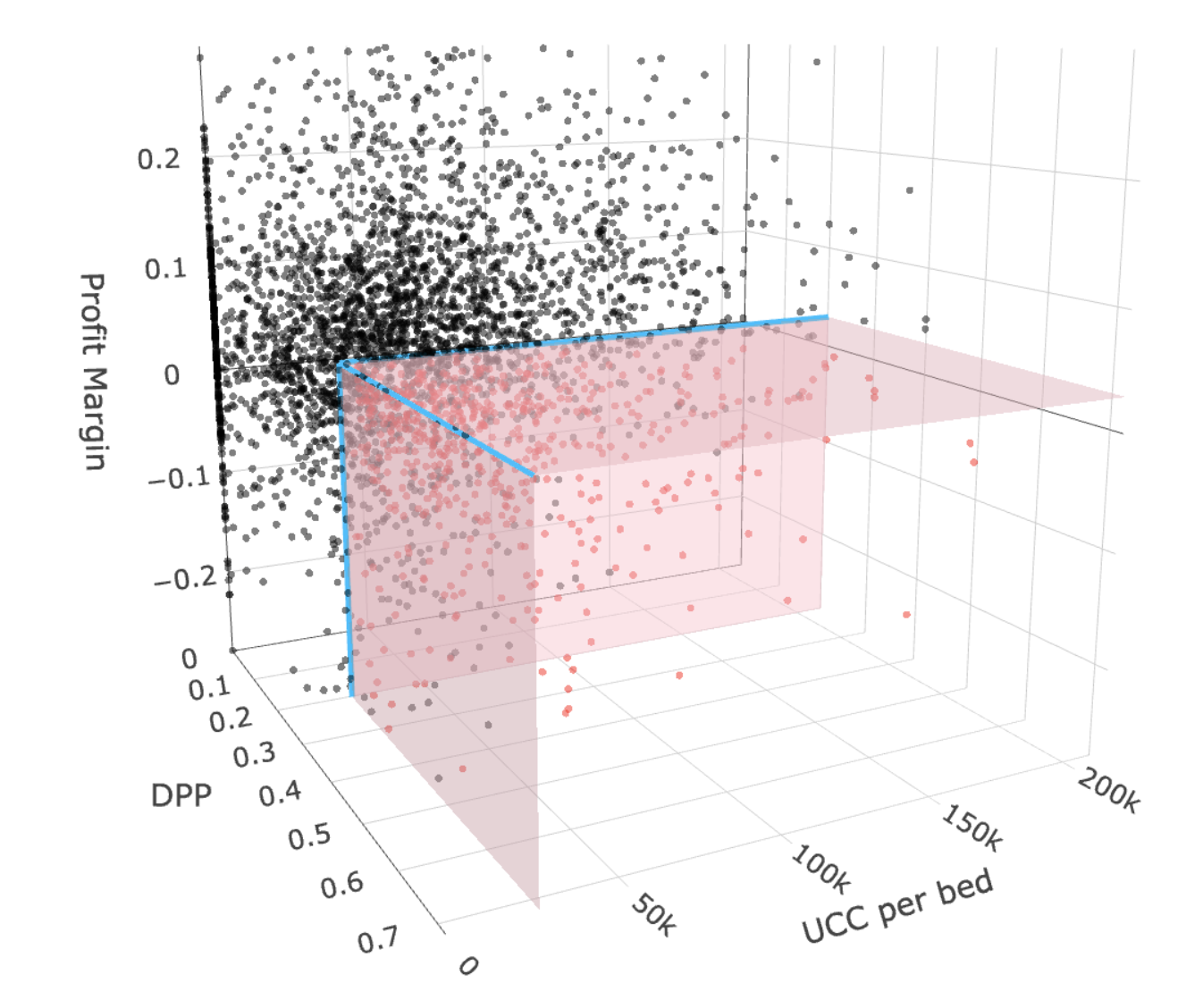}
   \caption{Visualization of the CARES Act data. The treatment boundary is shown in pink, with its edges colored in blue to represent variation along a single running variable while holding the other two at their respective cutoffs. The sample observations are displayed as a scatter plot, with dots colored in black (red) denoting control (treated) observations.}
   \label{fig:empvis}
\end{figure}

We follow the steps in Appendix H of \cite{NY23} for data cleaning and missing value imputation.\footnote{In addition, we remove one hospital that appears twice in the outcome data, as the two occurrences have considerably different values for the outcome variable.} The resulting treated and control sample sizes are $n_1 = 656$ and $n_0 = 3$,199, respectively. Figure \ref{fig:empvis} visualizes the sample data as a scatter plot, where red (black) colored dots correspond to treated (control) hospitals; the treatment boundary is consisted of three partially bounded planes colored in pink. Under the assumptions in Section \ref{sec:setup}, we can identify the conditional treatment effect of funding eligibility at any point on the pink surface. 

\subsection{Empirical Results}
Safety net funding eligibility criteria create a three-dimensional RD setup, where our framework applies. Acknowledging that directly estimating the conditional treatment effects nonparametrically at each treatment boundary point $X_i$ ``is hard to use in practice, however, when $X_i$ has many elements (p. 12)," \citet{NY23} propose a 2SLS estimator that is consistent for a weighted average of conditional treatment effects, nesting both propensity score matching and regression discontinuity in the case where the overlap condition fails in the first stage. In their empirical application, they use funding eligibility as an instrument for the amount of funding received and estimate the average effect of received funding amount on multiple outcomes, including the number of COVID patients hospitalized. Their findings suggest little to no effect. We find a similar null effect, but our analysis views funding eligibility as a binary treatment and directly estimates its causal effect. Moreover, we estimate conditional treatment effects as a function of treatment boundary point, allowing investigation of treatment effect heterogeneity along the treatment boundary. We find the estimated treatment effects, albeit with confidence intervals (CIs) covering zero, exhibit sizable heterogeneity along the treatment boundary.

We apply the honest regression forest and local linear forest estimators in \eqref{eqn:rf-estimator} and \eqref{eqn:llf-estimator} to the CARES Act data and, for ease of illustration, we plot how the treatment effect varies along each of the three running variables, while holding the other two fixed. This corresponds to estimating the treatment effect along the three ``edges" of the treatment boundary, shown in blue in Figure \ref{fig:empvis}. Due to the potential zero-density issue of local linear regressions with a collapsed univariate score, and the inability of the minimax-optimal estimator to accommodate more than two scores, we omit analyses using these two estimators. 

\begin{figure}[H]
\centering
    \includegraphics[width=0.9\textwidth]{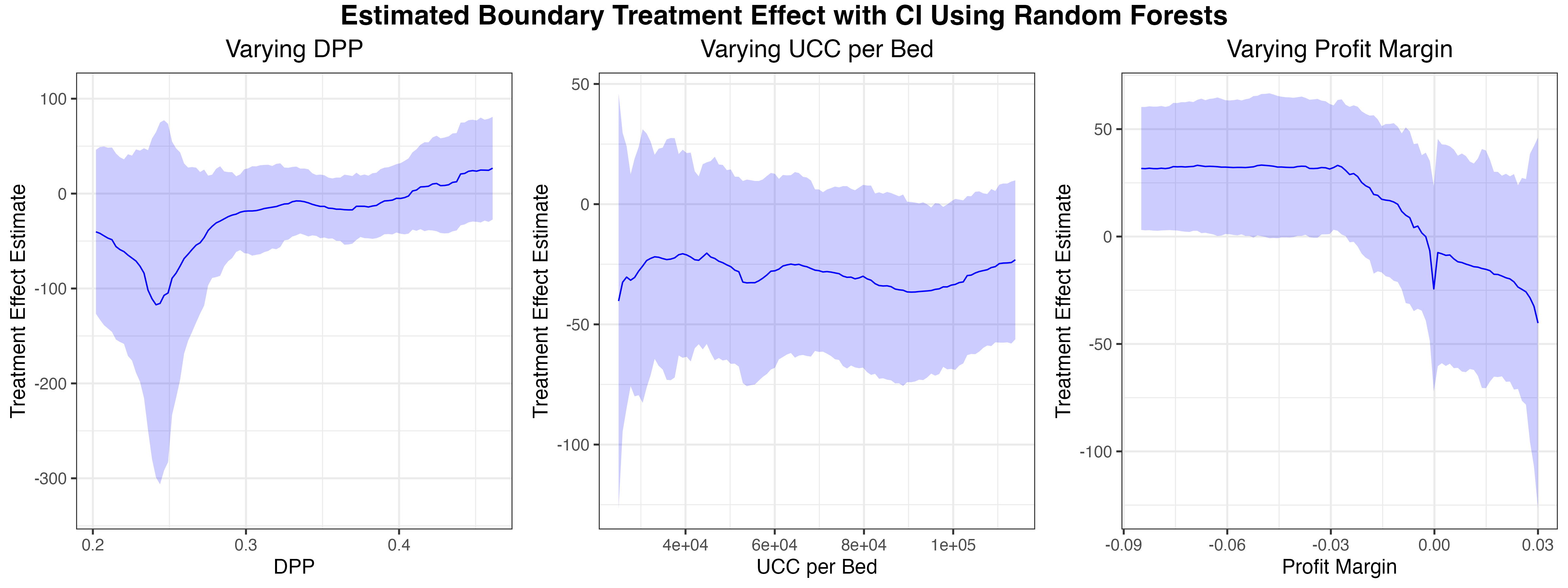}
    
    \vspace{0.25cm}
    \includegraphics[width=0.9\textwidth]{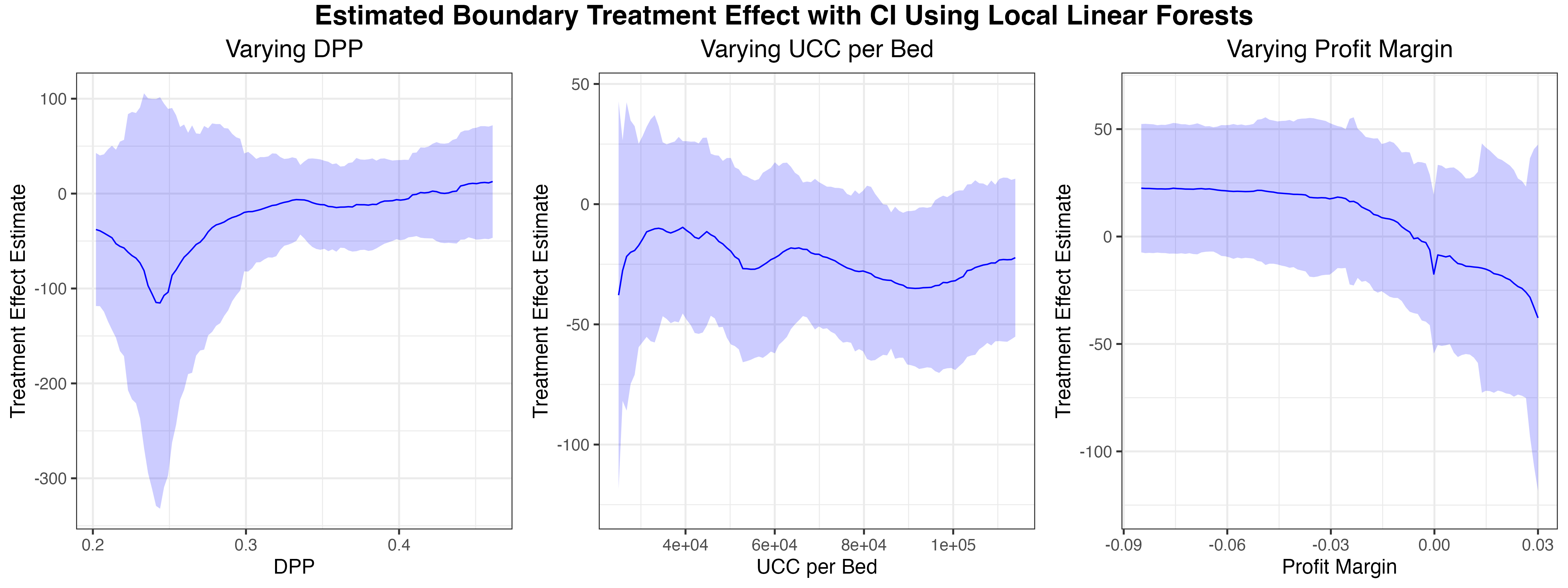}
   \caption{Treatment effect estimates (blue lines) with $95$\% CIs (blue shaded regions) for the honest regression forest estimator (top three panels) and the local linear forest estimator (bottom three panels).}
   \label{fig:emp}
\end{figure}    

For simplicity and better generalizability, we do not tune the hyperparameters of the forest estimators, but instead use the default values and set a few to specific pilot values as detailed in Section \ref{appendix:forest-param}. We use 50,000 trees for more stable results, as the \texttt{grf} package is known to be irreproducible across platforms.\footnote{See the \texttt{grf}
\href{https://grf-labs.github.io/grf/REFERENCE.html\#forests-predict-different-values-depending-on-the-platform-even-though-the-seed-is-the-same}{package manual}.} Figure \ref{fig:emp} presents the estimation results with 95\% CIs. Comparing across the top and bottom panels, we see little qualitative difference between the two estimators, and the CIs for the point estimates generally cover zero, supporting \cite{NY23}'s conclusion that the relief funding was not well targeted toward hospitals most in need. There is, however, noticeable treatment effect heterogeneity as DPP and profit margin vary, suggesting that more vulnerable hospitals---those with higher DPP and lower profit margin---tend to utilize the relief funding more efficiently by providing care to more patients.

\section{Conclusion}
\label{sec:end}
This paper studies RD designs with multiple scores. We first highlight the critical but often overlooked zero-density issue with running local linear regressions using a collapsed univariate score, and then explore the use of nonparametric estimators based on forests that directly model multivariate scores. Our simulation results suggest that the zero-density issue, while hard to detect from an empirical dataset, is likely present and would impair the performance of local linear regressions. Regression forests and local linear forests, on the other hand, generally require larger sample sizes to achieve good performance at boundary points, though local linear forests are better at capturing linear trends. Although this paper does not offer a simple, universal solution to multivariate RD problems, it adds two forest-based estimators to empirical researchers' toolkit. Overall, we recommend using the two forest-based methods as the primary approach when the sample size is large and the number of scores exceeds two because of their flexibility and theoretical validity. In cases where the sample size is small, local linear regressions with a collapsed univariate score can serve as a starting point, but researchers should proceed with caution and perhaps compare the results against other nonparametric alternatives introduced in this paper as a way of robustness checks.

\newpage
\bibliographystyle{chicago}
\bibliography{draft_LQ}

\newpage
\singlespacing
\bigskip
\begin{center}
{\large\bf SUPPLEMENTARY MATERIAL}
\end{center}

\appendix
\spacingset{1.45}
\setcounter{figure}{0}
\renewcommand\thefigure{\thesection.\arabic{figure}}
\setcounter{table}{0}
\renewcommand\thetable{\thesection.\arabic{table}}
\section{Proof}
\label{appendix:proof}
\noindent\textbf{Proof of Proposition \hyperref[prop:1]{1}.} Define $Z=(X_2-c_2)^2$. Consider the mapping $(X_1,Z)\mapsto(E,V)$, which is one-to-one. We can find the joint density of $(X_1,Z)$ by
\begin{align*}
    \mathbb{P}(X_1\leq x_1,Z \leq z) &=  \mathbb{P}\left(X_1\leq x_1, c_2-\sqrt{z}\leq X_2 \leq c_2+\sqrt{z} \right)\\
    &=F_{(X_1,X_2)}\left(x_1,c_2+\sqrt{z}\right)-F_{(X_1,X_2)}\left(x_1,c_2-\sqrt{z}\right)\\
    \implies f_{(X_1,Z)}(x_1,z)&=\frac{\partial^2{\mathbb{P}(X_1\leq x_1,Z \leq z)}}{\partial x_1\partial z}\\
    &=\frac{1}{2\sqrt{z}}\left[f_{(X_1,X_2)}\left(x_1,c_2+\sqrt{z}\right)+f_{(X_1,X_2)}\left(x_1,c_2-\sqrt{z}\right)\right].
\end{align*}
Since $X_1=V+c_1$ and $Z=E^2-V^2$, we have the Jacobian
\begin{align*}
    J_{(X_1,Z)}(e,v)=\begin{bmatrix}
        0&1\\
        2e&-2v
    \end{bmatrix}
\end{align*}
and $\lvert J_{(X_1,Z)}(e,v) \rvert=2e$. By the bivariate transformation theorem, 
\begin{align*}
    f_{(E,V)}(e,v)&=f_{(X,Z)}(x,z)\lvert J_{(X_1,Z)}(e,v) \rvert\\
    &=\frac{e}{\sqrt{e^2-v^2}}\left[f_{(X_1,X_2)}\left(v+c_1,c_2+\sqrt{e^2-v^2}\right)+f_{(X_1,X_2)}\left(v+c_1,c_2-\sqrt{e^2-v^2}\right)\right].
\end{align*}
This implies that the marginal distribution of $E$ is 
\begin{align*}
    f_E(e)=\int_{l(e)}^{u(e)}\frac{e}{\sqrt{e^2-v^2}}\left[f_{(X_1,X_2)}(v+c_1,\sqrt{e^2-v^2}+c_2)+f_{(X_1,X_2)}(v+c_1,-\sqrt{e^2-v^2}+c_2)\right]dv
\end{align*}
and the claim follows.\hfill $\square$

\subsection{Examples}
\label{appendix:eg}
\noindent\textbf{Example A.1.} Let $(X_1,X_2)$ be uniformly distributed over $[-1,1]^2$. Then $f_{(X_1,X_2)}(x_1,x_2)=\frac{1}{4}$ over the support $\Omega(X_1,X_2)=[-1,1]^2$. First consider the boundary point at $(0,0)$. The bivariate transform $(E,V)$ in this case has joint support 
\begin{align*}
    \Omega(E,V)& =\left\{(e,v)\in \mathbb{R}^2: e\in [0,\sqrt{2}]\right., \\
    &\left.\qquad\qquad v \in [-e,e] \text{ if } e \in [0,1] \text{ and } v \in \left[\sqrt{e^2-1},1\right]\cup \left[-1,-\sqrt{e^2-1}\right] \text{ if } e \in \left(1,\sqrt{2}\right]\right\}.
\end{align*}
Then 
\begin{align*}
    f_E(e)&=\mathbf{1}\{e\in[0,1]\}\int_{-e}^e\frac{e}{2\sqrt{e^2-v^2}}dv \\
    &\qquad\qquad\qquad\qquad+ \mathbf{1}\{e\in(1,\sqrt{2}]\}\left(\int_{\sqrt{e^2-1}}^{1}\frac{e}{2\sqrt{e^2-v^2}}dv+\int_{-1}^{-\sqrt{e^2-1}}\frac{e}{2\sqrt{e^2-v^2}}dv\right).
\end{align*}
Following Proposition \hyperref[prop:1]{1},
\begin{numcases}{f_E(e)=}
    \pi e/2, & \text{if } $e\in [0,1]$\notag\\
    e\left[\sin^{-1}(1/e)-\sin^{-1}(\sqrt{e^2-1}/e)\right], & \text{if } $e\in (1,\sqrt{2}]$\label{unif}.
\end{numcases}
Hence $f_E(0)=0$. Note that this is not specific to the chosen boundary point, $(0,0)$. In fact, the issue remains for any boundary point $(c_1,c_2)\in [-1,1]^2$ with respect to which we calculate the Euclidean distance. To see this, suppose without loss of generality that $c_1,c_2>0$; other cases follow similarly. Then 
\begin{align*}
    E=\sqrt{(X_1-c_1)^2+(X_2-c_2)^2},\quad V=X_1-c_1,
\end{align*}
where
\begin{align*}
    |V|\leq E \leq \sqrt{V^2+(1+c_2)^2},\quad E\in\left[0,\sqrt{(1+c_1)^2+(1+c_2)^2}\right].
\end{align*}
Then the joint support of $(E,V)$ when $E$ takes value in $[0,1-c_1]$ is
\begin{align*}
    \Omega\left(E,V \,|\, E\in [0,1-c_1]\right)=\left\{(e,v)\in\mathbb{R}^2: e\in [0,1-c_1], v\in[-e,e]\right\}.
\end{align*}
The case where $E\in\left(1-c_1, \sqrt{(1+c_1)^2+(1+c_2)^2}\right]$ is irrelevant to deriving $f_E(0)$. Then Equation (\ref{unif}) still holds for $e\in [0,1-c_1]$ and therefore $f_E(0)=0$.

\vspace{.5cm}

\noindent\textbf{Example A.2.} Let $(X_1,X_2)$ be independently normally distributed with mean $0$ and variance $\sigma^2$. Then $f_{(X_1,X_2)}(x_1,x_2)=\frac{1}{2\pi \sigma^2}\exp\{-x_1^2/2\sigma^2\}\exp\{-x_2^2/2\sigma^2\}$ over the support $\Omega(X_1,X_2)=\mathbb{R}^2$. First consider the boundary point at $(0,0)$. The bivariate transform $(E,V)$ in this case has joint support
\begin{align}
    \Omega(E,V)& =\left\{(e,v)\in \mathbb{R}^2: e\geq 0, v\in[-e,e]\right\}.\label{normal}
\end{align}
Then 
\begin{align*}
    f_E(e)&=\int_{-e}^e\frac{e}{\sqrt{e^2-v^2}}\frac{1}{\pi \sigma^2}\exp\left\{-e^2/2\sigma^2\right\}dv\\
    &=\frac{e}{\sigma^2}\exp\{-e^2/2\sigma^2\}
\end{align*}
Hence $f_E(0)=0$. Note that this is again not specific to the choice of the boundary point, as for any $(c_1,c_2)\in\mathbb{R}^2$, Equation (\ref{normal}) still holds.

\section{More on Simulations}
\label{appendix:supp}
This section provides additional information on the Monte Carlo simulations in Section \ref{sec:simulation}.

\subsection{Details on the Polynomial DGPs}
\label{appendix:polyDGP}
\subsubsection{Univariate Design: \citet{L08} Data}
Following \citet{IK12}, \citet{CCT14} and \citet{CCFT19}, we build the polynomial DGP using the \citet{L08} data as follows:
    \begin{align*}
        Y_i = \mu(X_i) +\epsilon_i, \quad X_i \sim (2{Beta}(2,4)-1), \quad \epsilon \sim N(0,\sigma^2)
    \end{align*}
where ${Beta}$ denotes the Beta distribution, $\mu(\cdot)$ is constructed by fitting a $5$-th order global polynomial with different coefficients for the treated and untreated groups taking the form:
\begin{align}\label{DGP:Lee}
  \mu(x) =
  \begin{cases}
  0.48+1.27x+7.18x^2+20.21x^3+21.54x^4+7.33x^5 & \text{if $x<0$} \\
  0.52+0.84x-3.00x^2+7.99x^3-9.01x^4+3.56x^5 & \text{if $x\geq0$}
  \end{cases}    
\end{align}
and $\sigma$ is estimated from the residuals of the regression and is equal to $0.1295$. Under this DGP, the true treatment effect at the $0$ cutoff is $0.04$.

\subsubsection{Bivariate Design: \citet{KT15} Data}
\label{sec:poly DGP}
\begin{figure}[H]
    \centering
    \includegraphics[width=1\textwidth]{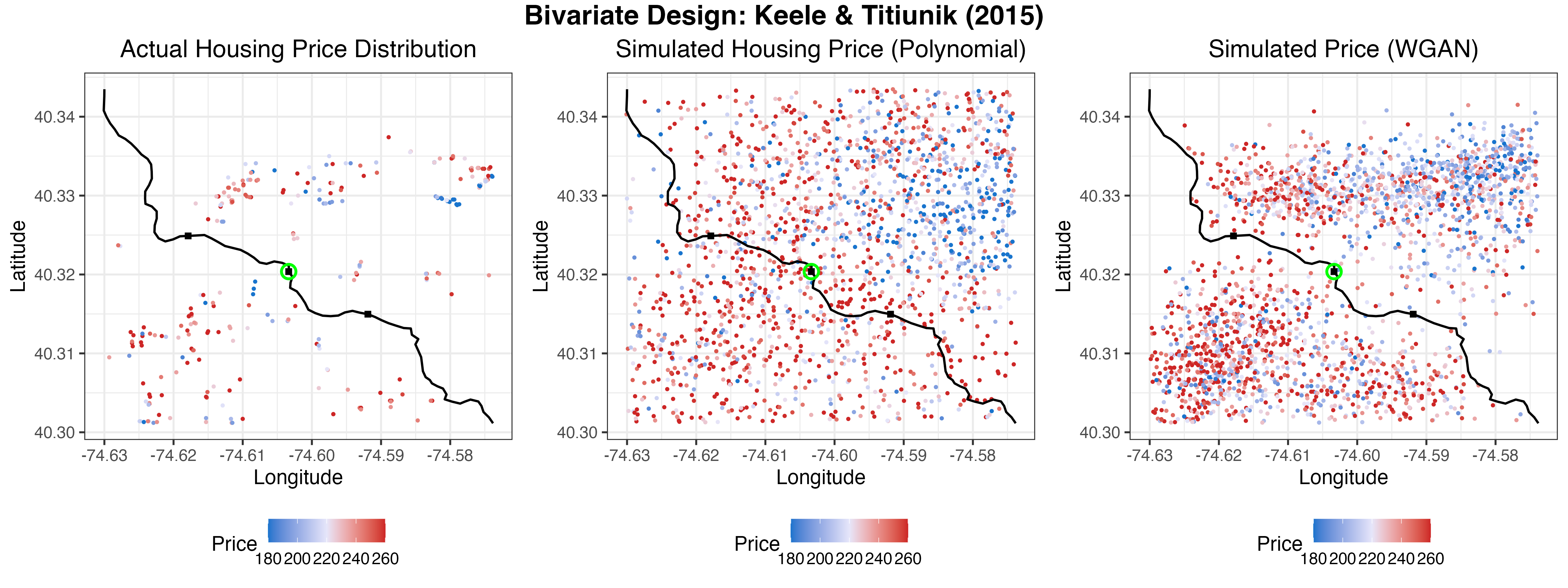}
    \vskip0.2cm
    \includegraphics[width=1\textwidth]{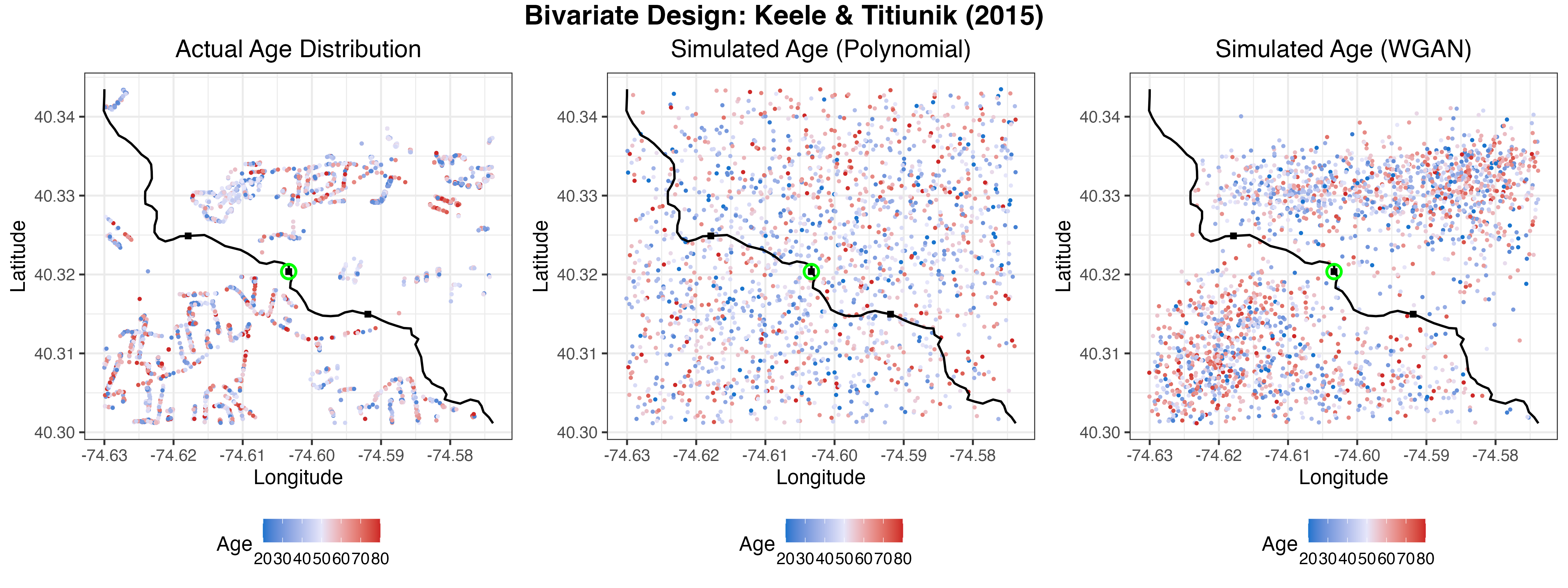}
    \vspace{-0.75cm}
    \caption{Visualization of the sample data in \citet{KT15} (left in each row) as well as simulated polynomial and WGAN DGPs (middle and right in each row, respectively). The outcome variable for the first (resp. second) row is housing price (resp. age). Coordinates below the black curve (treatment boundary) are treated, and the green circle shows the boundary point at which we estimate the treatment effect. The plot for actual housing prices uses a different dataset, which has fewer observations than the other two outcome cases. To visualize the polynomial DGPs, 2,500 coordinates are sampled with replacement from the original sample plus a noise term $N(0,0.01^2)$. We fit $3$-rd order interacted polynomials for housing price and age, and the plots show the model predictions at each pair of coordinates sampled.}
    \label{fig:KT-price&age}
\end{figure}

For the continuous outcome variables, housing price and age, the models are of the form 
\[{Y_i}_j = \mu_j(X_i) +\epsilon_{ij}, \text{ where } j \in \{\text{price}, \text{age}\} \text{ and } \epsilon_{ij} \sim N(0,\sigma_j^2).\]
We fit third-order interacted polynomials for $\mu_j(X_i)$. Below are the model specifications, where $\Tilde{x}_{i1}, \Tilde{x}_{i2}$ are demeaned versions of $x_{i1}, x_{i2}$ (i.e., latitude, longitude) respectively: 
\[\mu_{price}(x_i) =
  \begin{cases}
  13544.3+150.2{x}_{i1}-11847.6\Tilde{x}_{i1}^2-29821.7\Tilde{x}_{i1}^3\\
  \quad+259.3{x}_{i2} -15479.8\Tilde{x}_{i2}^2-207469.6\Tilde{x}_{i2}^3 +24446.9\Tilde{x}_{i1}\Tilde{x}_{i2}& \text{if $i$ is treated,} \\
  230407.6-9713.5{x}_{i1}+537644.5\Tilde{x}_{i1}^2-8356369.6\Tilde{x}_{i1}^3\\
  \quad-2164.6{x}_{i2} -9998.4\Tilde{x}_{i2}^2+748352.9\Tilde{x}_{i2}^3 +55631.1\Tilde{x}_{i1}\Tilde{x}_{i2}& \text{if $i$ is control,}
  \end{cases}
\]
and $\sigma_{price} = 32.6334$.
\[\mu_{age}(x_i) =
  \begin{cases}
  -477.8-70.0{x}_{i1}-111.9\Tilde{x}_{i1}^2-54704.1\Tilde{x}_{i1}^3\\
  \quad-44.9{x}_{i2} +4564.6\Tilde{x}_{i2}^2+107699.6\Tilde{x}_{i2}^3 -5863.3\Tilde{x}_{i1}\Tilde{x}_{i2}& \text{if $i$ is treated,} \\
  77307.5-1026.2{x}_{i1}+60847.1\Tilde{x}_{i1}^2-749930.1\Tilde{x}_{i1}^3\\
  \quad+481.1{x}_{i2} -13224.5\Tilde{x}_{i2}^2+185693.5\Tilde{x}_{i2}^3 -16725.5\Tilde{x}_{i1}\Tilde{x}_{i2}& \text{if $i$ is control,}
  \end{cases}
\]
and $\sigma_{age} = 15.9496$. For the discrete outcome, election turnout, we consider a logit conditional expectation function $P(Y_i = 1 | X_i) = \frac{e^{poly(X_i)}}{1+e^{poly(X_i)}}$, where $poly(X_i)$ is a third-order interacted polynomial as before. Below is the specification for $poly(X_i)$: 
\[poly(x_i) =
  \begin{cases}
  -200.4+3.7{x}_{i1}+224.4\Tilde{x}_{i1}^2+1028.1\Tilde{x}_{i1}^3\\
  \quad-0.7{x}_{i2} +57.1\Tilde{x}_{i2}^2+3778.1\Tilde{x}_{i2}^3 -127.5\Tilde{x}_{i1}\Tilde{x}_{i2}& \text{if $i$ is treated,} \\
  10399.6-286.5{x}_{i1}+10516.0\Tilde{x}_{i1}^2-114884.8\Tilde{x}_{i1}^3\\
  \quad-15.4{x}_{i2} -423.8\Tilde{x}_{i2}^2+12700.1\Tilde{x}_{i2}^3 +228.7\Tilde{x}_{i1}\Tilde{x}_{i2}& \text{if $i$ is control.}
  \end{cases}
\]
Figure \ref{fig:KT-price&age} visualizes the polynomial and WGAN DGPs for age and housing price.

\subsection{Estimating the Second Derivative Bound}
\label{appendix:sec-deriv-bound}
In this section, we consider estimating the second derivative bound $\mathcal{B}$ for the univariate polynomial DGP based on the real data from \cite{L08} by fitting a global quadratic function and multiplying the maximal estimated curvature by different constants. The results are reported in Table \ref{tbl:lee-B}, along with the results using the infeasible least upper bound $\mathcal{B}$ calculated from the true polynomial DGP in \eqref{DGP:Lee}. \cite{IW19} recommend multiplying by $2$ or $4$, but in this example these choices do not yield reasonable estimates for the second derivative bound and thus lead to biased estimates and poor coverage rates, as in Panels A-B, Columns 6-10 of Table \ref{tbl:mc-results}.  Only after multiplying the maximal estimated curvature by $30$ do we get a $\hat{\mathcal{B}}$ close to the true $\mathcal{B}$.

\begin{table}[H] \centering 
\resizebox{\columnwidth}{!}{
\begin{tabular}
{@{\extracolsep{5pt}}lccccccccc} 
\hline
\vspace{-.28cm}
\\
 & \multicolumn{3}{c}{True $\mathcal{B}=14.36$}
 & \multicolumn{3}{c}{$\hat{\mathcal{B}}=2\times$max.curvature} & \multicolumn{3}{c}{$\hat{\mathcal{B}}=4\times$max.curvature}\\
\cline{2-4} \cline{5-7} \cline{8-10}
\vspace{-.2cm}
\\
\multicolumn{1}{r}{Sample size $n=$} & $1,000$ & $5,000$ & $10,000$ & $1,000$ & $5,000$ & $10,000$ & $1,000$ & $5,000$ & $10,000$
\vspace{.2cm}
\\[-1.2ex]
\hline \\[-1.2ex] 
\multicolumn{1}{l}{Bias} & 0.0538 & 0.0471 & 0.0453 & 0.0432 & 0.0345 & 0.0297 & 0.0364 & 0.0252 & 0.0211\\
\multicolumn{1}{l}{Variance} & 0.0019 & 0.0006 & 0.0003 & 0.0006 & 0.0002 & 0.0001 & 0.0009 & 0.0002 & 0.0002\\
\multicolumn{1}{l}{95\% coverage} & 96.7\% & 96.7\% & 97.0\% & 67.6\% & 36.7\% & 24.7\% & 80.3\% & 69.7\%	& 64.0\%\\
\multicolumn{1}{l}{Average $\hat{\mathcal{B}}$} & - & - & - & 0.8372 & 0.8363 & 0.8369 & 1.6555 & 1.6736 & 1.6791\\
\\[-1.2ex]\hline 
\vspace{-.28cm}
\\
 & \multicolumn{3}{c}{$\hat{\mathcal{B}}=6\times$max.curvature}
 & \multicolumn{3}{c}{$\hat{\mathcal{B}}=10\times$max.curvature} & \multicolumn{3}{c}{$\hat{\mathcal{B}}=30\times$max.curvature}\\
\cline{2-4} \cline{5-7} \cline{8-10}
\vspace{-.2cm}
\\
\multicolumn{1}{r}{Sample size $n=$} & $1,000$ & $5,000$ & $10,000$ & $1,000$ & $5,000$ & $10,000$ & $1,000$ & $5,000$ & $10,000$
\vspace{.2cm}
\\[-1.2ex]
\hline \\[-1.2ex] 
\multicolumn{1}{l}{Bias} & 0.0306 & 0.0203 & 0.0166 & 0.0241 & 0.0145 & 0.0120 & 0.0115 & 0.0067 & 0.0056\\
\multicolumn{1}{l}{Variance} & 0.0011 & 0.0003 & 0.0002 & 0.0013 & 0.0004 & 0.0002 & 0.0020 & 0.0005 & 0.0003\\
\multicolumn{1}{l}{95\% coverage} & 87.6\% & 85.0\% & 80.3\% & 93.0\% & 90.1\% & 89.8\% & 95.9\%	& 96.5\% & 96.4\%\\
\multicolumn{1}{l}{Average $\hat{\mathcal{B}}$} & 2.5116 & 2.5161 & 2.5217 & 4.1695 & 4.1910 & 4.1817 & 12.6209 & 12.5831 & 12.5993\\
\\[-1.2ex]\hline 
\end{tabular}
}
\\
\caption{Simulation results based on different estimates of the second derivative bound. The true least upper bound on the second derivative of the DGP (\ref{DGP:Lee}) is $\mathcal{B}=14.36$. The table reports the average bias, variance, and $95\%$ coverage rate over the 1,000 Monte Carlo replications using the true bound as well as estimated bounds by fitting a global quadratic and then multiplying the maximal curvature of the fitted model by different constants $\in\{2,4,6,10,30\}$. Also reported is the Monte Carlo average of estimated bounds.}
\label{tbl:lee-B}
\end{table} 

\subsection{Details on the WGAN DGPs}
\label{appendix:wganDGP}
To construct artificial datasets that closely resemble the observed samples in \citet{KT15} near the treatment boundary, we employ the WGAN approach in \citet{AIMM21} and constrain the input data to a window defined by the start and end points of the treatment boundary, i.e., using the boundary as a rough diagonal. This window and the sample data points involved are represented by the plots on the left in Figures \ref{fig:KT-turnout} and \ref{fig:KT-price&age}. The resulting input data consists of 2,628 treated and 2,969 control observations for turnout and age, and 117 treated and 206 control observations for housing price.

\begin{figure}[H]
    \centering
    \subfigure{
        \includegraphics[width=0.229\textwidth]{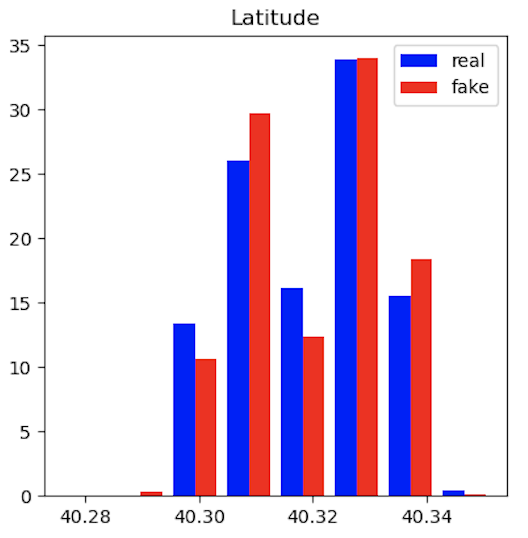}}
    \subfigure{
        \includegraphics[width=0.235\textwidth]{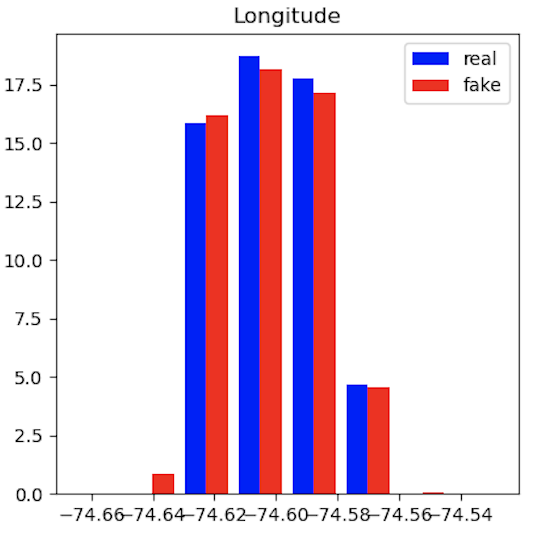}}
    \subfigure{
        \includegraphics[width=0.229\textwidth]{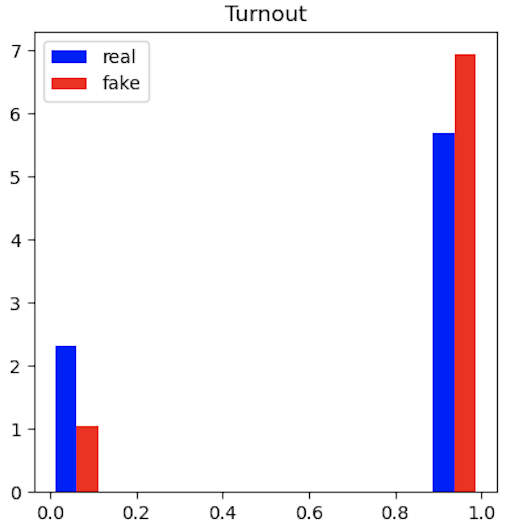}}
    \subfigure{
        \includegraphics[width=0.247\textwidth]{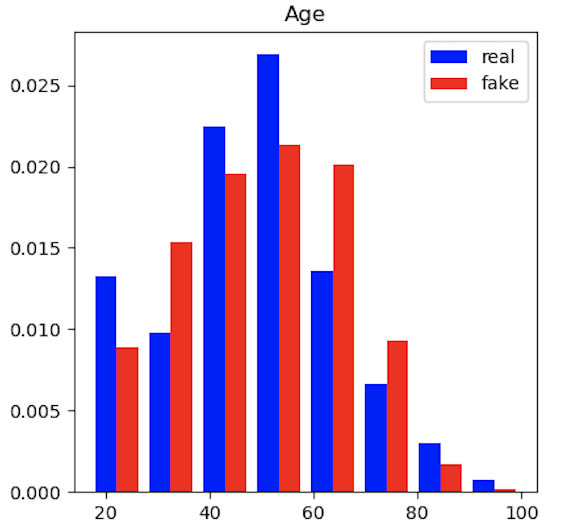}}
    \caption{Marginal histograms for the actual turnout and age samples (real) and their WGAN-generated counterparts (fake). The two observed samples share the same geographic coordinates, so the latitude and longitude comparisons are produced once (two plots on the left).}
    \label{fig:WGAN turnout age hist}
\end{figure}

\begin{figure}[H]
    \centering
    \subfigure{
        \includegraphics[width=0.24\textwidth]{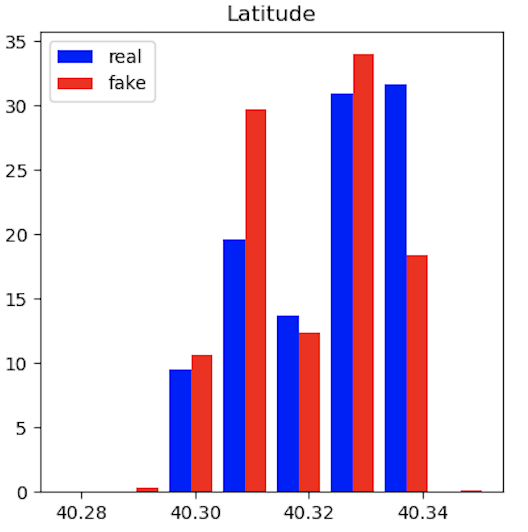}}
    \subfigure{
        \includegraphics[width=0.244\textwidth]{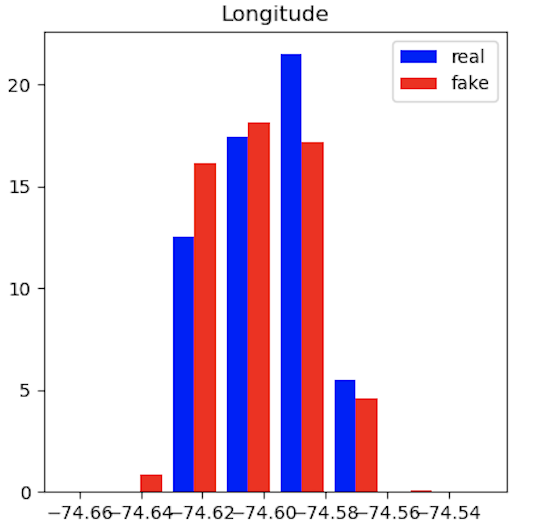}}
        \subfigure{
        \includegraphics[width=0.255\textwidth]{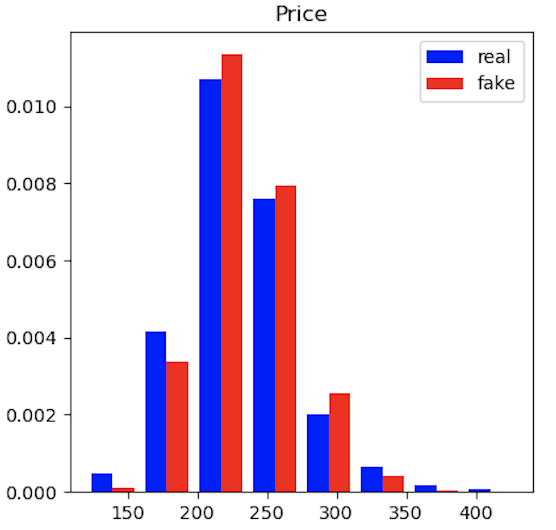}}
    \caption{Marginal histograms for the actual housing price sample (real) and its WGAN-generated counterpart (fake). The geographic coordinates in the observed price sample are a subset of those in the turnout/age sample, so the latitude and longitude comparisons are regenerated, while the WGAN coordinates are generated once and fixed across the three outcomes.}
    \label{fig:WGAN price hist}
\end{figure}

\begin{figure}[H]
    \centering
    \includegraphics[width=0.5\textwidth]{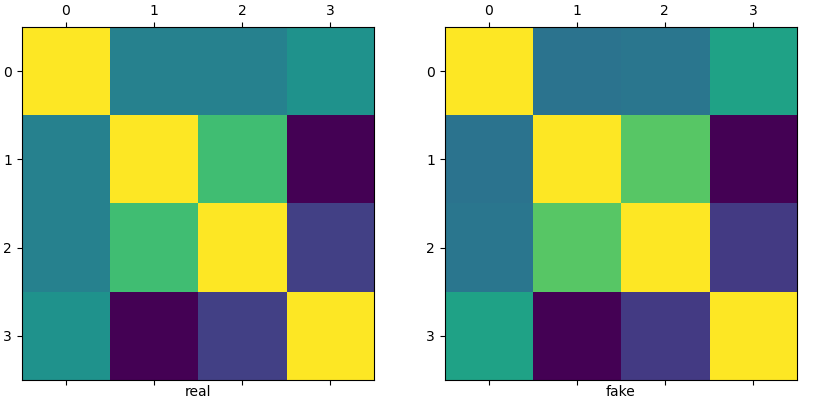}
    \caption{Between-variable correlations for the observed turnout sample (real) and its WGAN-generated counterpart (fake).}
    \label{fig:WGAN turnout corr}
\end{figure}

\begin{figure}[H]
    \centering
    \includegraphics[width=0.5\textwidth]{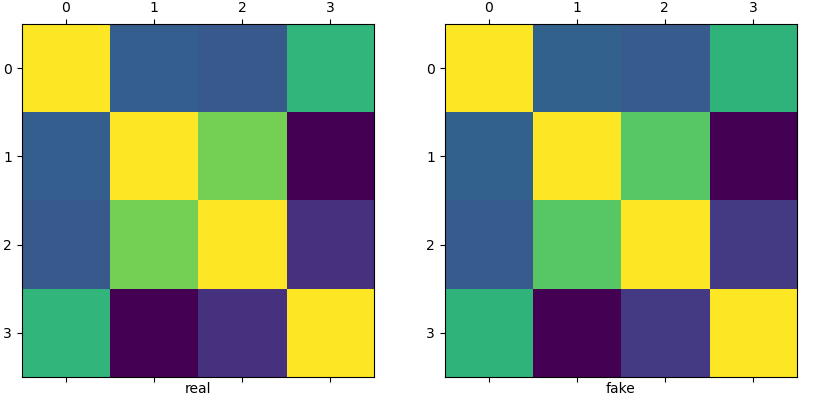}
    \caption{Between-variable correlations for the observed housing price sample (real) and its WGAN-generated counterpart (fake).}
    \label{fig:WGAN price corr}
\end{figure}

\begin{figure}[H]
    \centering
    \includegraphics[width=0.5\textwidth]{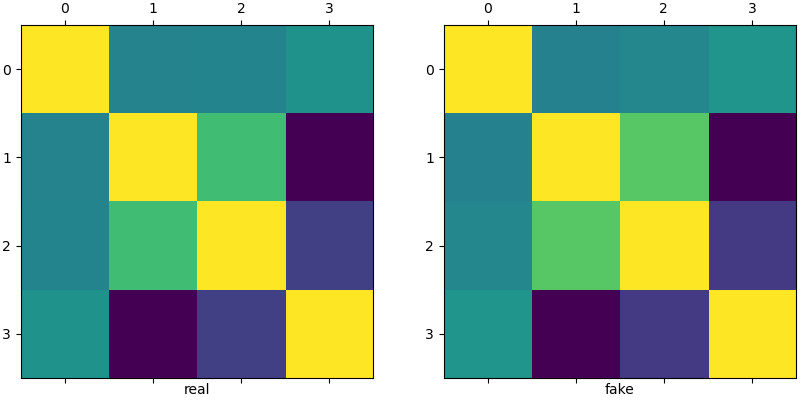}
    \caption{Between-variable correlations for the observed age sample (real) and its WGAN-generated counterpart (fake).}
    \label{fig:WGAN age corr}
\end{figure}

We use the default options available from the \texttt{wgan} package of \citet{AIMM21} to train the WGAN for the results reported in Table \ref{tbl:mc-results}.\footnote{These default specifications can be found at \texttt{wgan} \href{https://ds-wgan.readthedocs.io/en/latest/api.html\#specifications}{package manual}.} We note that the current implementation of WGAN is only available in \texttt{Python}, whereas implementations of the minimax-optimal estimator and the two forests estimators are in \texttt{R}, and therefore we cannot, for each Monte Carlo replication, generate a new sample from WGAN in \texttt{Python}, keep the generated sample in memory, and then take it to \texttt{R} and fit the estimators: for one iteration of $1000$ Monte Carlo replications and each with a sample size of $n$, such procedure requires $n\times1000$ observations, which quickly runs out of computer memory as we increase $n$. Instead, for each dataset that we fit a WGAN to, we draw a WGAN-generated sample of $40$ million observations, and use this large artificial sample to run the Monte Carlo simulations, treating it as a ``superpopulation" from which we draw random subsamples of size $n$ used in each Monte Carlo replication.\footnote{Each WGAN sample is corrected based on the geographic coordinates of the treatment boundary to ensure that for each hypothetical household, the generated coordinates match the generated treatment status. This correction is minor, affecting fewer than 2\% of the observations in the WGAN sample.} Figures \ref{fig:WGAN turnout age hist}--\ref{fig:WGAN age corr} display graphical comparisons between the actual turnout, housing price and age samples and their WGAN-generated counterparts. 

\subsection{Wasserstein Distances between Real and Artificial Data}
\citet{AIMM21} combine generative adversarial networks (GANs) with the Wasserstein distance objective to create a hypothetical dataset that closely mimics the observed sample, enabling effective Monte Carlo simulations. In this subsection, we consider the two-dimensional example in \citet{KT15} and show that the datasets constructed by WGAN are indeed more similar to the actual samples in terms of Wasserstein distance, as compared to the datasets generated by the polynomial DGPs in Section \ref{sec:poly DGP}. While we use the default options when training the WGAN DGPs used in the main text (training the neural network for 1,000 epochs, for example), we report the Wasserstein distances for 2,000, 3,000, and 4,000 epochs as well. As elaborated in Section \ref{sec:robust epoch}, these specifications serve as robustness checks for the estimators proposed in the main text. 

Following \citet{AIMM21}, we compute the exact Wasserstein distances via linear programming, averaging over 10 random samples from the WGAN dataset, each of equal size to the original dataset. Table \ref{tbl:wd} organizes the Wasserstein distances for the outcome variables considered in Section \ref{appendix:wganDGP}. We note that the Wasserstein distance between distributions $\mathbb{P}$ and $\mathbb{P'}$, defined as
\[W(\mathbb{P}, \mathbb{P'}):=\inf_{\gamma\in\Pi(\mathbb{P}, \mathbb{P'})}\mathbb{E}_{(R,S)\stackrel{d}{\sim}\gamma}\left[||R-S||\right]\]
with $\Pi(\mathbb{P}, \mathbb{P'})$ denoting the set of joint distributions with marginals $\mathbb{P}$ and $\mathbb{P'}$ such that $R\stackrel{d}{\sim}\mathbb{P}$ and $S\stackrel{d}{\sim}\mathbb{P'}$, is not normalized with respect to the scale of random variables $R$ and $S$. Thus, its magnitude can depend on the scale of the data being compared. This explains why housing price has the largest Wasserstein distances and turnout has the smallest. Comparing across rows in Table \ref{tbl:wd}, all the WGAN DGPs considerably outperform the polynomial ones in terms of Wasserstein distance. Within the WGAN entries, the distances for turnout and age tend to shrink with increasing training epochs. However, for housing price, the small sample size (117 treated and 206 control observations) may explain the instability in training the networks and calculating the distances.

\begin{table}[H]
\centering
\footnotesize
\begin{tabular}{lccccc}
\toprule
& \multirow{2}{*}{Polynomial} & \multicolumn{4}{c}{WGAN} \\
\cmidrule(r){3-6}
& & 1,000 Epochs & 2,000 Epochs & 3,000 Epochs & 4,000 Epochs \\
\midrule
Turnout & 142.8712 & 0.1603 & 0.0147 & 0.0242 & 0.0047\\
Housing Price & 330.3712 & 3.6796 & 6.5035 & 6.1220 & 7.0828\\
Age & 147.0418 & 1.9824 & 1.1613 & 1.3046 & 1.0996\\
\bottomrule
\end{tabular}
\\
\caption{Wasserstein distances between generated and observed data.}
\label{tbl:wd}
\end{table}

\subsection{Robustness of the WGAN simulations}
\label{appendix:wganRobust}
For the bivariate designs in Table \ref{tbl:mc-results}, results on the housing price outcome appear to be most sensitive to changes in sample size and estimator choice. To complement Section \ref{sec:MCresults}, we consider three robustness checks: robustness to subsamples, model architectures, and the number of training epochs. The first two checks are also considered by \cite{AIMM21} to assess the ranking of a set of estimators for the average effect for the treated. We omit the third robustness check in \cite{AIMM21}---robustness to the size of training data---since the sample size of the housing price data (117 treated and 206 control observations) is too small for most of the fractions considered in \citet[Table 11]{AIMM21}. 

\subsubsection{Robustness to Subsamples}
Following \citet[Section 5.1]{AIMM21}, we run the WGAN algorithm on 10 subsamples randomly drawn without replacement from the original data; each subsample is 80\% of the size of the original data.\footnote{More specifically, the WGAN coordinates ($X_i$) and treatment statuses ($D_i$) are generated from 80\% of the turnout/age data, and the corresponding housing prices ($Y_i|(X_i,D_i)$) are generated from 80\% of the housing price data. This is because the coordinates in the housing price data are a subset of those in the turnout/age data, which has more observations.} For each subsample, we train a WGAN superpopulation of size 5 million and run 1,000 Monte Carlo simulations with sample size 10,000 for each of the four estimators to estimate the treatment effect at the boundary point.\footnote{We consider a smaller superpopulation size because we need to train a WGAN for each of the subsamples drawn, and repeatedly generating a superpopulation of size 40 million as in Section \ref{sec:simulation} turned out to be too computationally intensive.} Table \ref{tbl:robustness subsample} shows the bias, variance and coverage averaged across the 10 hypothetical populations trained from the 10 random subsamples of the original data. Standard deviations are enclosed in parentheses. The averaged metrics from using the subsamples do not differ considerably from those from using the full sample, see Table \ref{tbl:mc-results}, Panel G under $n=$ 10,000.

\begin{table}[H]
\centering
\scriptsize
\begin{tabular}{lcccc}
\toprule
 & Local Linear Regression & Minimax-Optimal\ & Honest Regression Forest\ & Honest Local Linear Forest \\
\midrule
Bias & 5.2632 (1.4239) & 0.0176 (0.5703) & 4.7418 (1.3558) & 3.5876 (0.9605) \\
Variance & 82.4645 (25.0978) & 106.13 (42.6781) & 28.7466 (6.2132) & 42.3938 (12.0164) \\
95\% coverage & 94.22\% (0.71\%) & 98.07\% (0.59\%) & 84.88\% (5.71\%) & 88.43\% (3.05\%) \\
\bottomrule
\end{tabular}
\\
\caption{Robustness results for housing price: average and standard deviations of metrics over 10 subsamples drawn from the original sample. Each random subsample for generating the WGAN superpopulation is 80\% of the original sample size. The sample size is 10,000 for each Monte Carlo trial.}
\label{tbl:robustness subsample}
\end{table}
\subsubsection{Robustness to Model Architectures}
The default architecture in the WGAN algorithm for the generator and critic neural networks consists of three hidden layers with 128 units each. To examine whether our qualitative comparisons in Section \ref{sec:MCresults} are sensitive to the architecture specification, we adopt two alternative model architectures from \citet[Section 5.2]{AIMM21}. Both alternatives have a three-layer generator and a three-layer critic. In the first alternative, the generator (critic) has layer dimensions of $[64, 128, 256]$ ($[256, 128, 64]$). In the second alternative, the generator (critic) has layer dimensions of $[128, 256, 64]$ ($[64, 256, 128]$). Both WGAN superpopulations are of size 5 million, and the sample size is 10,000 for each Monte Carlo replication. Results from varying the model architecture are organized in Table \ref{tbl:robustness architecture}. 

\begin{table}[H]
\centering
\scriptsize
\begin{tabular}{lcccc}
\toprule
 & Local Linear Regression & Minimax-Optimal\ & Honest Regression Forest\ & Honest Local Linear Forest \\
\\[-2ex]\hline 
\multicolumn{5}{l}{\cellcolor{blue!20}\textbf{Main (generator: $[128, 128, 128]$, critic: $[128, 128, 128]$)}}\\
Bias & 5.6543 & $-0.0903$ & 4.9030 & 4.2545 \\
Variance & 80.8610 & 109.4033 & 31.7393 & 44.7507 \\
95\% coverage & 93.0\% & 97.2\% & 85.3\% & 85.5\% \\
\\[-2ex]\hline 
\multicolumn{5}{l}{\cellcolor{blue!20}\textbf{Alternative 1 (generator: $[64, 128, 256]$, critic: $[256, 128, 64]$)}}\\
Bias & 4.6898 & 0.1933 & 2.9349 & 2.3774 \\
Variance & 60.4756 & 88.1598 & 23.5457 & 31.2027 \\
95\% coverage & 95.0\% & 97.8\% & 90.6\% & 91.4\% \\
\\[-2ex]\hline 
\multicolumn{5}{l}{\cellcolor{blue!20}\textbf{Alternative 2 (generator: $[128, 256, 64]$, critic: $[64, 256, 128]$)}}\\
Bias & 6.1217 & 0.3147 & 6.6993 & 5.5907 \\
Variance & 97.7364 & 155.9217 & 31.0610 & 48.2774 \\
95\% coverage & 93.3\% & 97.3\% & 78.4\% & 81.8\% \\
\bottomrule
\end{tabular}
\\
\caption{Robustness results for housing price, with the first (second) alternative architecture having a generator hidden layer with dimensions $[64, 128, 256]$ ($[128, 256, 64]$) and a critic hidden layer with dimensions $[256, 128, 64]$ ($[64, 256, 128]$). The sample size 10,000.}
\label{tbl:robustness architecture}
\end{table}

\subsubsection{Robustness to the Number of Training Epochs}
\label{sec:robust epoch}
\begin{figure}[H]
    \centering
    \includegraphics[width=0.65\textwidth]{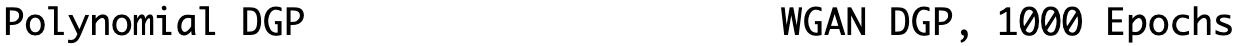}
    \includegraphics[width=0.9\textwidth]{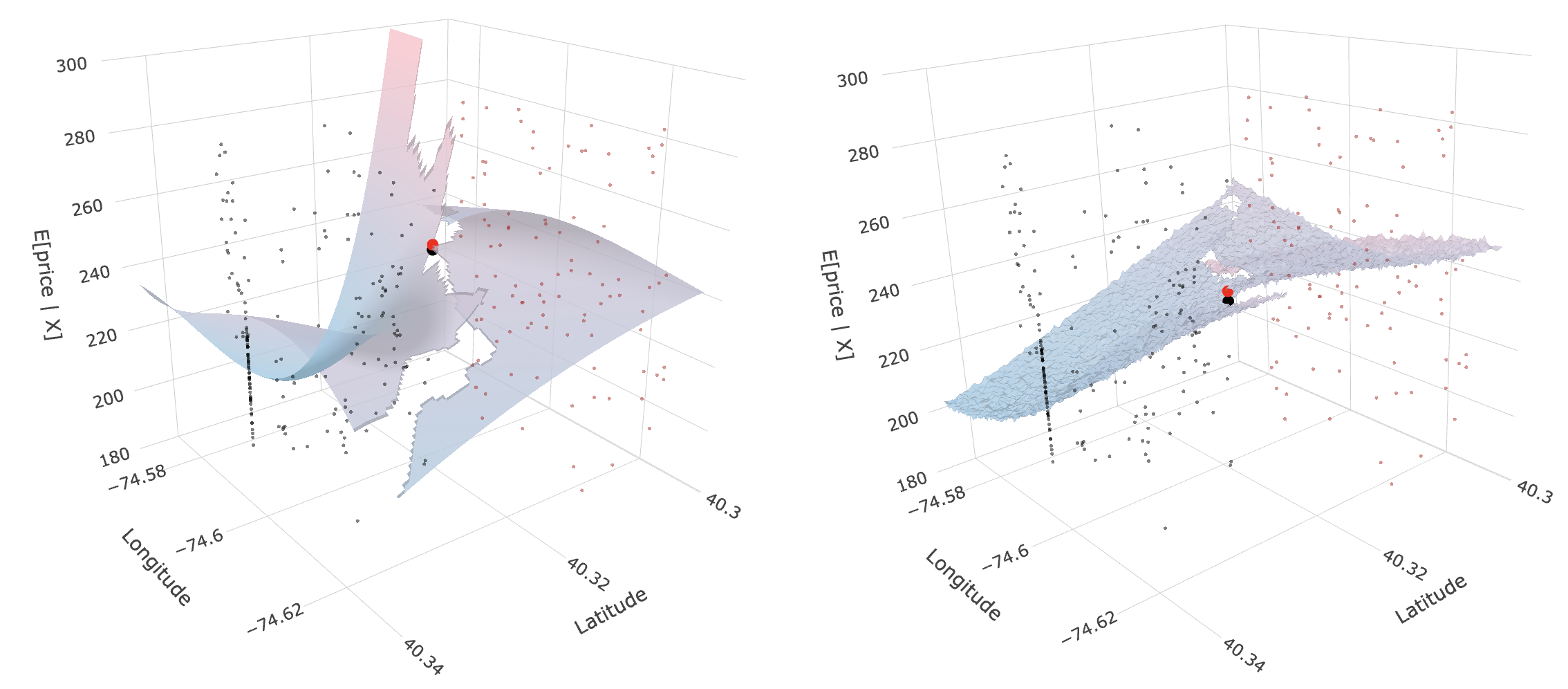}
    \includegraphics[width=0.09\textwidth]{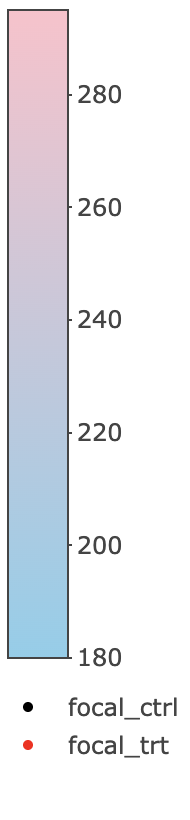}
    \includegraphics[width=0.925\textwidth]{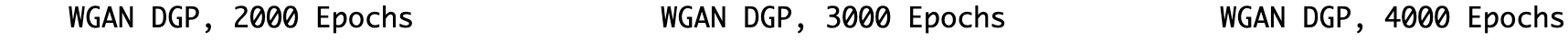}
    \includegraphics[width=1\textwidth]{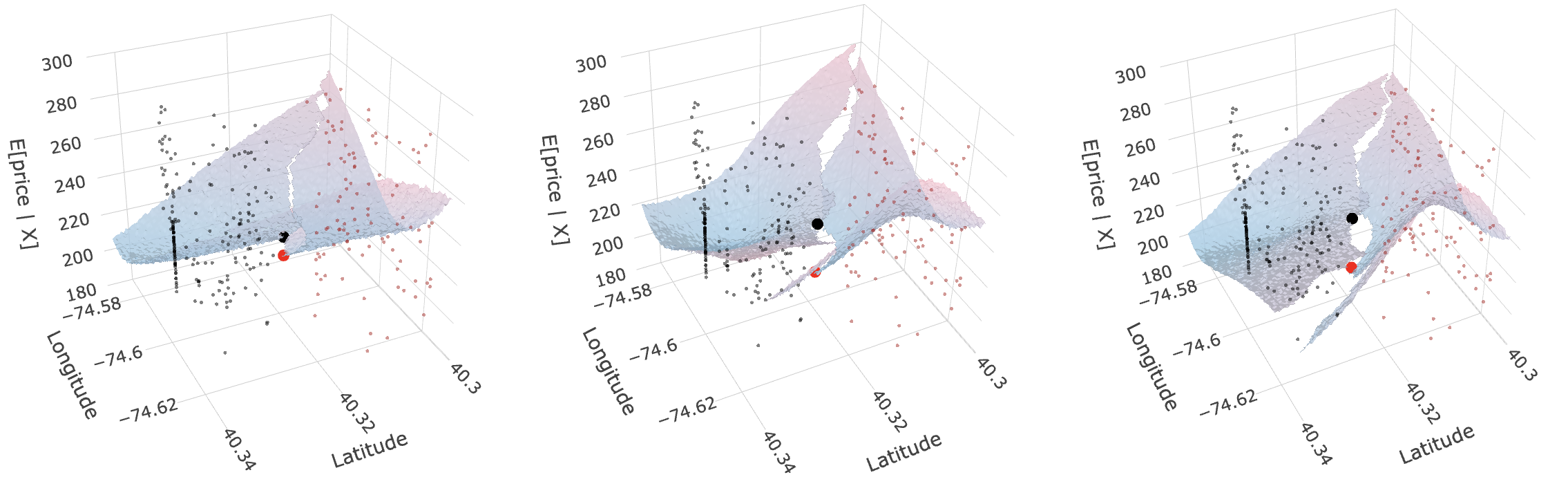}
    \vspace{-0.5cm}
    \caption{Surface plots for simulated housing price using the polynomial DGP detailed in Section \ref{sec:poly DGP} and the WGAN DGP introduced in Section \ref{sec:buildDGP} with 1,000, 2,000, 3,000 and 4,000 training epochs. Each point on the surfaces represents the average of 2,000 simulated housing prices at the corresponding geographic location. The outcome at the focal point is evaluated both as treated (solid red dot) and not treated (solid black dot), also averaged over 2,000 simulations. The original sample is displayed as a scatter plot around the surfaces, with red (treated) and black (control) semi-transparent dots.}
    \label{fig:surface main}
\end{figure}

In this subsection, we vary the number of training epochs specified in the WGAN algorithm for the housing price outcome as a robustness check. We consider running the algorithm for 2,000, 3,000, and 4,000 epochs, noting that 1,000 is the default option used for constructing the datasets in the main text. Each Monte Carlo trial uses a sample size of 10,000, which is moderate-to-large for RD studies. Figure \ref{fig:surface main} shows three-dimensional surface plots that visualize the simulated CEFs of housing price from the polynomial and the four WGAN DGPs, with the original observations scattered around the surfaces (i.e., CEFs). While the WGAN surfaces generally exhibit less curvature than the polynomial ones, we point out that the curvature of the treated WGAN surface (surrounded by treated dots in red) tends to increase with the number of training epochs. This instability may be attributed to the small sample size of the original data, which includes only 117 treated observations. Nevertheless, such variability in curvature could highlight certain strengths and weaknesses of the four estimators discussed in the main text. Table \ref{tbl:robustness epoch} presents the simulation results using the three alternative DGPs. Similar to the 1,000-epoch DGP in the main text, for $n=$ 10,000, local linear regressions perform well, and the minimax-optimal estimator overcovers.\footnote{Although not shown in Table \ref{tbl:robustness epoch}, we note that consistent with Panel G in Table \ref{tbl:mc-results}, the coverage rates for local linear regressions again deteriorate as we further increase the sample size. This deterioration is possibly due to the zero-density issue of the collapsed multivariate score.} Honest regression forests, however, gradually fail to capture linear signals in the treated surface near the boundary point as the surface becomes more tilted. Local linear forests alleviate this problem through the inclusion of a linear correction, as shown by their robust performance across DGPs. Nevertheless, forest-based estimators generally require larger sample sizes to achieve good coverage rates.

\begin{table}[H]
\centering
\scriptsize
\begin{tabular}{lcccc}
\toprule
 & Local Linear Regression & Minimax-Optimal\ & Honest Regression Forest\ & Honest Local Linear Forest \\
\\[-2ex]\hline 
\multicolumn{5}{l}{\cellcolor{blue!20}\textbf{2,000 Epochs}}\\
Bias & 2.2475 & $-0.8902$ & 6.8735 & 3.4928 \\
Variance & 190.9314 & 142.4837 & 24.5066 & 71.0147 \\
95\% coverage & 94.3\% & 99.4\% & 73.8\% & 85.0\% \\
\\[-2ex]\hline 
\multicolumn{5}{l}{\cellcolor{blue!20}\textbf{3,000 Epochs}}\\
Bias & $-3.2831$ & $-1.8100$ & 10.7442 & 4.4552 \\
Variance & 304.8468 & 122.9466 & 21.3419 & 82.1174 \\
95\% coverage & 93.9\% & 99.3\% & 40.5\% & 79.9\% \\
\\[-2ex]\hline 
\multicolumn{5}{l}{\cellcolor{blue!20}\textbf{4,000 Epochs}}\\
Bias & $-3.4504$ & $-1.2560$ & 11.3784 & 3.1766 \\
Variance & 258.7163 & 151.0802 & 24.7439 & 103.9739 \\
95\% coverage & 93.7\% & 98.4\% & 39.0\% & 81.4\% \\
\bottomrule
\end{tabular}
\\
\caption{Robustness results for housing price, with \{2,000, 3,000, 4000\} training epochs in the WGAN specification. The sample size 10,000.}
\label{tbl:robustness epoch}
\end{table}

\subsection{Hyperparameters for Regression and Local Linear Forests}
\label{appendix:forest-param}
For all forest-based estimators, we set the number of trees (argument \texttt{B} of the \texttt{regression\_forest()} function in the \texttt{grf} package) to 5,000 for the simulations and 50,000 for the CARES Act empirical application. More trees generally reduce overfitting but at a higher computational cost. At each sufficiently large tree node, we randomly pick one feature to make the split ($\texttt{mtry}=1$). Note that in univariate designs, $\texttt{mtry}$ is always 1. Small values of $\texttt{mtry}$ can %lead to trees with high variance but low bias and 
reduce the between-tree correlations, which would lower the overall variance. For regression forests, our choice of the subsample fraction is based on the condition in \citet[Theorem 5]{ATW19} that the subsample sizes $s_1$ and $s_0$, for the treated and control samples respectively,  should satisfy $s_1 \asymp n_1^{\beta}\,\,\,\text{and} \,\,\, s_0 \asymp n_0^{\beta}$ for some $\beta$ lower bounded by $\beta_{min}^{RF}:=\left(1+ \frac{\texttt{mtry}\times\log(1-\alpha)}{d\log(\alpha)}\right)^{-1}$, where $\alpha$ is the maximum imbalance of a split that defaults to $0.05$. Given this condition, we set the subsample fraction to $ceil[n_j^{\beta_{min}^{RF}}]/n_j$ times a constant $c\in[0.05,0.5]$ for $j \in \{0,1\}$. We leave the other tunable parameters of \texttt{regression\_forest()} at their default values for simplicity and generalizability.\footnote{See the \texttt{grf} \href{https://cran.r-project.org/web/packages/grf/grf.pdf}{package manual}.} 

\begin{figure}[H]
    \centering
    \includegraphics[width=0.85\textwidth]{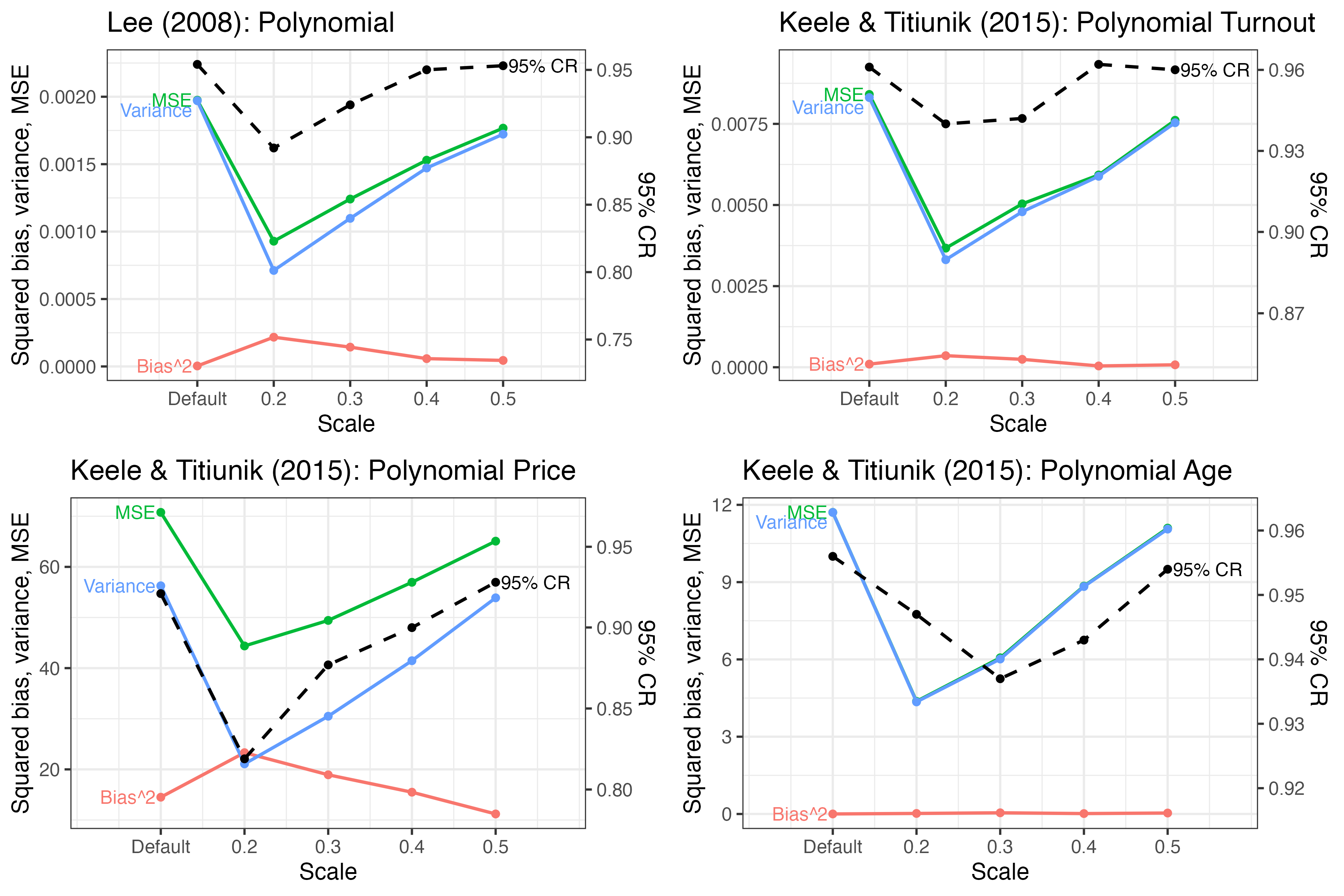}
    \vspace{-0.25cm}
    \caption{Comparisons of different choices of the scaling constant $c$ using the datasets simulated by the polynomial DGPs. The horizontal axis corresponds to $c \in \{0.2, 0.3, 0.4, 0.5\}$ as well as the “Default” option, in which case the subsample fraction defaults to $0.5$ without depending on $c$, $\beta$, or $n$. The left vertical axis corresponds to values of the Monte Carlo bias squared (red), variance (blue), and MSE (green); the right vertical axis corresponds to values of the empirical $95\%$ coverage rate (black).}
    \label{fig:scaling-poly}
\end{figure}

\begin{figure}[H]
    \centering
    \includegraphics[width=0.85\textwidth]{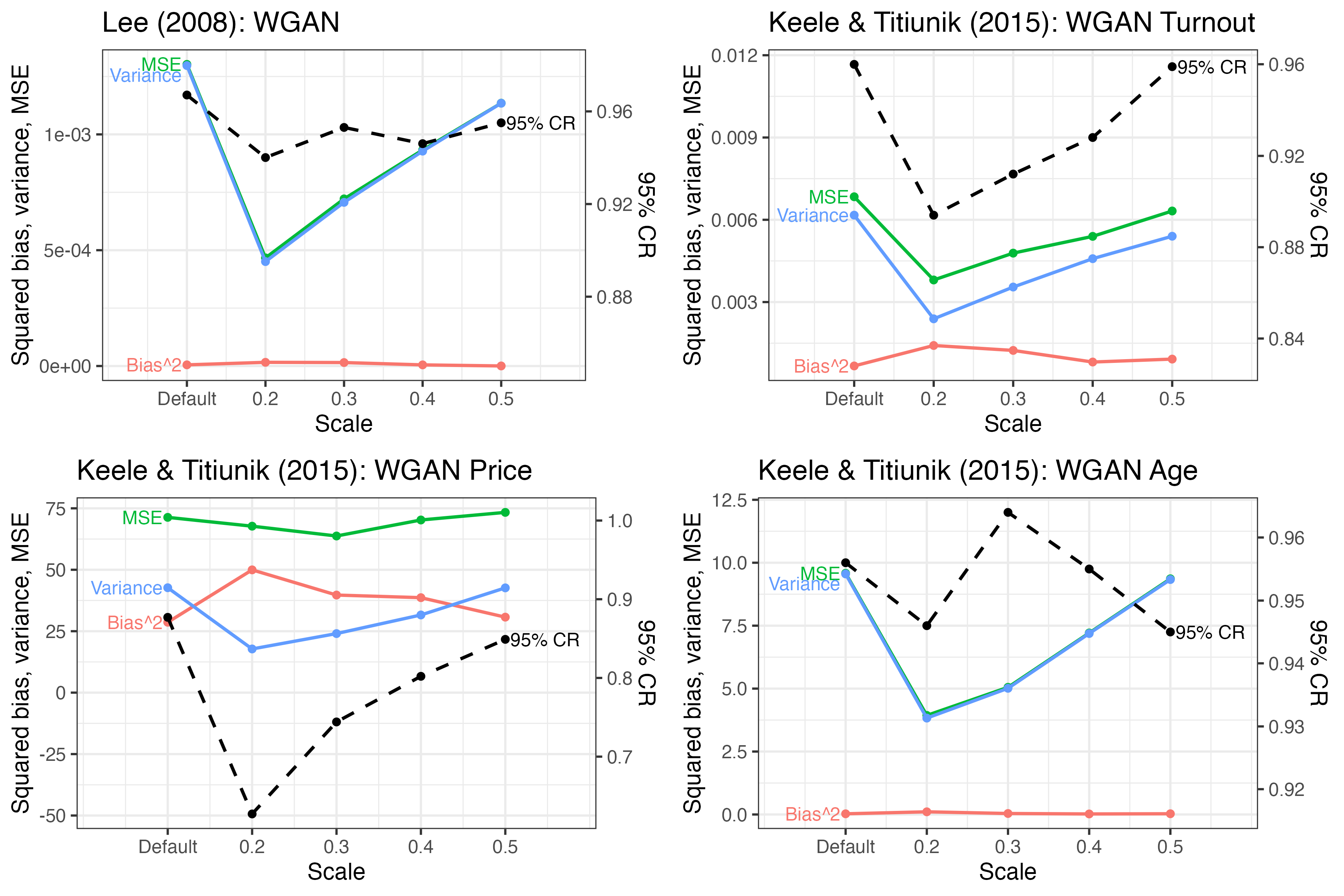}
    \vspace{-0.25cm}
    \caption{Comparisons of different choices of the scaling constant $c$ using the datasets simulated by the WGAN DGPs. The horizontal axis corresponds to $c \in \{0.2, 0.3, 0.4, 0.5\}$ as well as the “Default” option, in which case the subsample fraction defaults to $0.5$ without depending on $c$, $\beta$, or $n$. The left vertical axis corresponds to values of the Monte Carlo bias squared (red), variance (blue), and MSE (green); the right vertical axis corresponds to values of the empirical $95\%$ coverage rate (black).}
    \label{fig:scaling-wgan}
\end{figure}

Figures \ref{fig:scaling-poly} and \ref{fig:scaling-wgan} show, for each DGP we consider with sample size $n=$ 5,000, how the squared bias, variance, MSE and the $95\%$ coverage rate across the 1000 Monte Carlo replications change with the scaling constant $c$ we use. As $c$ gets smaller, the bias problem and undercoverage become more pronounced, whereas the variance decreases. We recommend setting $c$ in the range of $[0.4,0.5]$ for small to moderate datasets; smaller values of $c$ can be considered with larger sample sizes. In this paper, the empirical application and simulations for all forest-based estimators are conducted with $c=0.4$.

For local linear forests, the lower bound for $\beta$ is instead given by \citet[Theorem 5]{FTAW20}: $\beta_{min}^{LLF}:=1-\left(1+\frac{d}{1.3\times \texttt{mtry}}\cdot\frac{\log(\alpha)}{\log(1-\alpha)}\right)^{-1}$. Following \cite{{FTAW20}}, we make residual splits (\texttt{enable.ll.split = TRUE}) to minimize the corresponding prediction errors on ridge regression residuals, as opposed to standard splits that minimize prediction errors from employing leaf-wide averages. We standardize the ridge penalty by the covariance matrix (\texttt{ll.split.weight.penalty = TRUE}). Other tunable parameters of \texttt{ll\_regression\_forest()} are set to their default values, except the ridge penalty term $\lambda$ is automatically tuned by cross-validation.

Although we do not tune the parameters in \texttt{regression\_forest()} and \texttt{ll\_regression\_forest()} for ease of computation and to demonstrate the general applicability of the default specifications, we note that tuning may improve performance in terms of MSE but worsen coverage rates. This occurs because tuning targets out-of-bag MSE and trades off between bias and variance, which can lead to more biased results but smaller variances, thereby affecting the coverage rates.
\end{document}